\documentclass[12pt,nofootinbib,preprint,tightenlines
]{revtex4}

\def\AnswerYes{y}
\def\ShowLabelsVersion{n}         
\def\ShowChangesVersion{n}        
\def\ShowAnnotationsVersion{n}    

\usepackage{graphicx}
\usepackage{amsmath}
\usepackage{bm} 
\usepackage{slashed}
\usepackage{xspace}

\newcommand{\beq}{\begin{eqnarray}}
\newcommand{\eeq}{\end{eqnarray}}
\newcommand{\bqa}{\begin{eqnarray}}
\newcommand{\eqa}{\end{eqnarray}}

\newcommand{\lsim}{\hspace*{0.2em}\raisebox{0.5ex}{$<$}
    \hspace{-0.8em}\raisebox{-0.3em}{$\sim$}\hspace*{0.2em}}

\newcommand{\ii}{\mathrm{i}}
\newcommand{\dd}{\mathrm{d}}

\renewcommand{\Im}{\mathrm{Im}}
\renewcommand{\Re}{\mathrm{Re}}

\newcommand{\w}{\ensuremath{\omega}}

\newcommand{\wlab}{\ensuremath{\omega_\mathrm{lab}}}

\newcommand{\thetalab}{\ensuremath{\theta_\mathrm{lab}}}

\newcommand{\alphae}{\ensuremath{\alpha_{E1}}}
\newcommand{\betam}{\ensuremath{\beta_{M1}}}
\newcommand{\gammaee}{\ensuremath{\gamma_{E1E1}}}
\newcommand{\gammamm}{\ensuremath{\gamma_{M1M1}}}
\newcommand{\gammaem}{\ensuremath{\gamma_{E1M2}}}
\newcommand{\gammame}{\ensuremath{\gamma_{M1E2}}}

\newcommand{\alphaep}{\ensuremath{\alpha_{E1}}}
\newcommand{\betamp}{\ensuremath{\beta_{M1}}}
\newcommand{\gammaeep}{\ensuremath{\gamma_{E1E1}}}
\newcommand{\gammammp}{\ensuremath{\gamma_{M1M1}}}
\newcommand{\gammaemp}{\ensuremath{\gamma_{E1M2}}}
\newcommand{\gammamep}{\ensuremath{\gamma_{M1E2}}}
\newcommand{\gammazerop}{\ensuremath{\gamma_{0}}}
\newcommand{\gammapip}{\ensuremath{\gamma_{\pi}}}

\newcommand{\muv}{\ensuremath{\mu^{(\mathrm{v})}}}

\newcommand{\kappas}{\ensuremath{\kappa^{(\mathrm{s})}}}
\newcommand{\kappav}{\ensuremath{\kappa^{(\mathrm{v})}}}

\newcommand{\piN}{\pi\mathrm{N}}
\newcommand{\gammaN}{\gamma \mathrm{N}}
\newcommand{\gammaonp}{\gamma\mathrm{p}}
\newcommand{\p}{\mathrm{p}}

\newcommand{\MN}{\ensuremath{M_\mathrm{N}}} 
\newcommand{\Mp}{\ensuremath{M_\mathrm{p}}} 
\newcommand{\MDelta}{\ensuremath{M_\Delta}} 
\newcommand{\DeltaM}{\ensuremath{\Delta_{\scriptscriptstyle M}}} 
\newcommand{\mpi}{\ensuremath{m_\pi}}     
\newcommand{\fpi}{\ensuremath{f_\pi}}

\newcommand{\MeV}{\ensuremath{\mathrm{MeV}}}
\newcommand{\fm}{\ensuremath{\mathrm{fm}}}
\newcommand{\ChiEFT}{$\chi$EFT\xspace}

\newcommand{\NXLO}[1]{N\ensuremath{{}^{#1}}LO\xspace}

\renewcommand{\Re}{\ensuremath{\mathrm{Re}}}
\renewcommand{\Im}{\ensuremath{\mathrm{Im}}}

\renewcommand{\deg}{\ensuremath{^\circ}}
\newcommand{\OdL}{Olmos de Le\'on}

\newcommand{\calL}{\mathcal{L}}
\newcommand{\calO}{\mathcal{O}}

\newcommand{\ga}{g_{\scriptscriptstyle A}}
\newcommand{\gpiNN}{g_{\pi{\scriptscriptstyle\text{NN}}}}
\newcommand{\gpiNDelta}{g_{\pi{\scriptscriptstyle\text{N}\Delta}}}

\newcommand{\gE}{\ensuremath{g_{\scriptscriptstyle E}}}
\newcommand{\gM}{\ensuremath{g_{\scriptscriptstyle M}}}
\newcommand{\CDeltaN}{C_{\mathrm{N}\Delta} }
\newcommand{\be}{\begin{equation}}
\newcommand{\ee}{\end{equation}}
\newcommand{\bea}{\begin{eqnarray}}
\newcommand{\eea}{\end{eqnarray}}

\ifx\ShowLabelsVersion\AnswerYes
  \usepackage[scriptsize]{myshowkeys} 
\fi

\ifx\ShowAnnotationsVersion\AnswerYes
  \newcommand{\comment}[1]{{\scriptsize\sffamily\bfseries{#1}}}
  \newcommand{\margin}[1]{\marginpar{\scriptsize\sffamily\bfseries{#1}}}
\else
  \newcommand{\comment}[1]{}
  \newcommand{\margin}[1]{}
  
\fi

\ifx\ShowChangesVersion\AnswerYes
  \usepackage{ulem}
   
  \newcommand{\delete}[1]{\sout{#1}}            
  \renewcommand{\emph}[1]{\textit{#1}}           
\else

  \newcommand{\sout}[1]{}
  \newcommand{\xout}[1]{}

  \newcommand{\delete}[1]{}
\fi

\begin{document}


\preprint{INT-PUB-12-048}
%
\title{%
  Compton scattering from the proton in an effective field theory with
  explicit Delta degrees of freedom}
\author{J.~A.~McGovern}\email{judith.mcgovern@manchester.ac.uk}
\affiliation{School of Physics and Astronomy, The University of Manchester,
  Manchester M13 9PL, UK}
\author{D.~R.~Phillips}\email{phillips@phy.ohiou.edu} \affiliation{Department
  of Physics and Astronomy and Institute of Nuclear and Particle Physics,
  Ohio University, Athens, Ohio 45701, USA\footnote{Permanent address} and\\
  School of Physics and
  Astronomy, The University of Manchester, Manchester M13 9PL, UK}

\author{H.~W.~Grie\ss hammer}\email{hgrie@gwu.edu} \affiliation{Institute for
  Nuclear Studies, Department of Physics, The George Washington University,
  Washington DC 20052, USA\footnote{Permanent address} and\\
  J\"ulich
  Centre for Hadron Physics and
  Institut f\"ur Kernphysik (IKP-3),\\
  Forschungszentrum J\"ulich, D-52428 J\"ulich, Germany}

\begin{abstract}
We analyse the proton Compton-scattering differential cross section for photon energies up to
  $325\;\MeV$ using Chiral Effective Field Theory ($\chi$EFT) and extract new values for the
 electric and magnetic polarisabilities of the proton. Our approach builds in the key physics in two different regimes:
photon energies $\omega\lesssim\mpi$ (``low energy"), and the higher energies where the $\Delta(1232)$ resonance plays a key role. The Compton amplitude is
complete at
  \NXLO{4}, $\calO(e^2\delta^4)$, in the low-energy region, and
  at NLO, $\calO(e^2\delta^0)$, in the resonance region. Throughout, the 
  Delta-pole graphs are dressed with $\piN$ loops and 
  $\gammaN \Delta$ vertex corrections.  %
  A statistically consistent database of proton Compton experiments is used to constrain the free parameters in our amplitude: the $M1$
  $\gammaN \Delta$ transition strength $b_1$ (which is fixed in the resonance region) and the polarisabilities
$\alphaep$ and $\betamp$ (which are fixed from data below $170\;\MeV$). In order to  
  obtain a reasonable fit, we find it necessary to add the
  spin polarisability $\gammamm$ as a free parameter, even though 
  it is, strictly speaking, predicted in $\chi$EFT at the order to which we work.
  We show that the fit  is
  consistent with the Baldin sum rule, and then use that sum rule
 to constrain $\alphaep + \betamp$. In this way we obtain
  $\alphaep=[10.65\pm0.35(\text{stat})\pm0.2(\text{Baldin})
  \pm0.3(\text{theory})]\times10^{-4}\;\fm^3$
  and $\betamp =[3.15\mp0.35(\text{stat})\pm0.2(\text{Baldin})
  \mp0.3(\text{theory})]\times10^{-4}\;\fm^3$, with
  $\chi^2= 113.2$ for $135$ degrees of freedom. A detailed rationale for the
theoretical uncertainties assigned to this result is provided.
\end{abstract} 
\maketitle
%
\section{Introduction}
\label{sec:intro}

Compton scattering  from the proton, $\gamma\mathrm{p}\to\gamma\mathrm{p}$, at photon energies up to a
few hundred MeV, has proven an excellent tool to study the subtle interplay of the
effective low-energy degrees of freedom of hadrons, and of the symmetries
and dynamics which govern them; see recent reviews for
details~\cite{Report,Sc05,Gr12}.  It reveals the chiral symmetry of QCD and the pattern of
its breaking by probing pion-cloud effects, and the properties of the
$\Delta(1232)$ as the lowest nucleonic excitation. Since it tests the (real and
virtual) excitation spectrum of the target, probing the two-photon response of
a proton also complements the information available in the one-photon
response (e.g.~form factors).   

Since the earliest experiments on the proton~\cite{Ox58,Hy59,Go60,Be60,De61,Ba66a,Ba66b,Gr67,Pu67,Ba74,Ba75,Ge76}, a particular goal has been an extraction of the electric and magnetic polarisabilities, $\alphae$ and $\betam$, which parametrise the stiffness of the proton against deformation in uniform, static  electric and magnetic fields.  At photon energies $\wlab$ below 50~MeV the Compton cross section is completely dominated by the Born terms, with the Thomson cross section as the low-energy limit. The polarisabilities manifest themselves as corrections to this point-like cross section which grow as $\wlab^2$, but there is only a narrow range of energy before faster-varying terms contribute significantly. At $\wlab\sim150$~MeV non-analytic structure associated with pion photoproduction appears, and around the same energy the tail of the Delta resonance also becomes apparent.  In order to extract the polarisabilities, therefore, we need a reliable description of these effects.
In this paper we analyse Compton-scattering data in such a framework in order to extract $\alphae$ and $\betam$  in a model-independent way.
Our tool  is Chiral Effective Field Theory, \ChiEFT, the low-energy version of QCD~\cite{Weinberg79,GL82,Gasser:1983yg,JM91a,JM91b,BKM,Sc02,BM07,Be08, SS12}. In it, theoretical uncertainties can be estimated using its power counting, namely
a systematic expansion scheme for physical observables.
As the baryonic sector of Chiral Perturbation Theory ($\chi$PT) including nucleons
and the $\Delta(1232)$, it orders all interactions consistent with the
symmetries of QCD (and in particular the pattern of its chiral-symmetry
breaking) by an expansion in powers of ``low'' momentum scales in units of a
``high'' scale $\Lambda$ at which new degrees of freedom become
relevant. Calculating both tree and loop diagrams to a given order in this small parameter results in an
amplitude that is truncated at a certain, prescribed level of accuracy.

Indeed, Compton scattering provided one of the early successes of  $\chi$EFT
applied to nucleons. Bernard et al.~\cite{Be91,Be92} calculated the polarisabilities at leading one-loop order, without 
explicit $\Delta(1232)$ degrees of freedom, to be
\begin{equation} 
  \label{eq:BKM}
  \alphae^\mathrm{LO}=10\betam^\mathrm{LO}=
  \frac{10e^2\ga^2}{192\pi\mpi\fpi^2}=12.6\times10^{-4}\;\fm^3,
\end{equation}
in remarkably good agreement with both the extractions of these parameters extant at that time, and those obtained in subsequent experiments.
Furthermore  the cross sections obtained from that theory were in qualitative agreement with  the then-available data below about
$150\;\MeV$ \cite{BKM,Ba97}. Subsequently, a large number of experiments \cite{Fe91,Zi92,Ha93,MacG95,Mo96,Pe96,Hu97,Wi99,Blanpied,OdeL,Wolf,Ca02} have measured unpolarised
differential cross sections at photon
energies between about 60 and 400~MeV; see the thorough discussions in the
reviews~\cite{Report,Sc05,Gr12}.  However, we shall argue in this article that the
database remains quite sparse in important regions, in particular between 150 and
250 MeV. In addition, there are unresolved consistency issues between data
sets from different experiments. Therefore, plans to measure
Compton cross sections at MAMI and
HI$\gamma$S~\cite{Mi12,Mi09,Ho12} are very welcome.

Results obtained for $\alphae$ and $\betam$ using \ChiEFT may be compared to direct
determinations from fully dynamical lattice QCD which appear
imminent~\cite{Detmold:2006vu,De09,De10,Lujan:2011ue,Engelhardt:2011qq}. Such
simulations, in turn, take advantage of the fact that \ChiEFT reliably
predicts the strong dependence of the polarisabilities on the pion mass, so
that results can be extrapolated from unphysical quark masses. Thus, Compton
scattering provides an excellent example of how \ChiEFT serves as a bridge between
cross sections measured in experiments and the non-perturbative quark-gluon
dynamics underlying the physics of hadrons.

Apart from its relevance in Compton scattering, the magnetic polarisability $\betam$  also contributes the largest error on the two-photon-exchange contribution to the Lamb shift in muonic hydrogen \cite{Pachucki,Carlson:2011dz, Bi12}. And, while we focus here on the case of the proton, this study
also provides input to extractions of neutron polarisabilities from data on
elastic Compton scattering  on deuterium; see e.g.~Refs~\cite{Ph09,Gr12}. Small proton-neutron differences stem from chiral-symmetry-breaking pion-nucleon interactions, probing details of QCD. For example, Walker-Loud et al.~\cite{WalkerLoud:2012bg} recently found that the biggest uncertainty in theoretical determinations of the electromagnetic proton-neutron mass difference now comes from the contribution of
$\approx[0.47 \pm 0.47$]~MeV from this effect~\cite{Gr12}. However, such effects can be probed with confidence only if both experimental and theoretical uncertainties are well under control.

As already mentioned, there has been a significant increase in the data available on $\gamma \mathrm{p}$ scattering over the last twenty years. Over the same period the EFT description of this process has been refined in several ways. The
pion-nucleon-loop amplitudes are now known to one order higher~\cite{Be93, Ra93,
  McG01}, where two short-distance contributions to $\alphae$ and $\betam$
enter the Compton amplitude. The polarisabilities are then no longer predicted, but can be determined from low-energy Compton data, and the fit
quality indicates the extent to which \ChiEFT correctly captures the energy-dependence
of the Compton amplitudes. This has been done in Refs.~\cite{Be02,Be04}. In
another line of attack, the $\Delta(1232)$ was included as an explicit degree of
freedom, applying Lagrangians developed in Refs.~\cite{He96,He97,He98,Hi04}. This
allows for a description which applies from zero photon energy into the Delta
resonance region, and thus also for using data at intermediate energies for alternative extractions of the  polarisabilities~\cite{PP03,Hi04}.
In both these variants, baryons have traditionally been included as
non-relativistic degrees of freedom in a version called ``Heavy-Baryon Chiral
Perturbation Theory'' (HB$\chi$PT)~\cite{JM91a,JM91b,Be92}. A good qualitative reproduction of Compton data in this
  regime has also  been obtained in a $\chi$EFT calculation which does not
  invoke the heavy-baryon expansion~\cite{BL00,Fu03,LP09,LP10}, and the agreement between the cross sections with and without the heavy-baryon expansion is very good---even at leading one-loop order---once the polarisabilities are fixed \cite{Le12}.   In line
with the power-counting philosophy, it is encouraging that the extractions or
predictions of $\alphae$ and $\betam$ have not varied dramatically from the
original predictions, as demonstrated by the values
advocated in a recent review~\cite{Gr12}. 

In this paper, we merge these three refinements: higher orders in the chiral
sector, explicit Delta degrees of freedom, and partially-covariant
formulations. In $\chi$EFT  one identifies two low-energy scales, $\mpi$ and $\MDelta-M_N$.  Different  power countings involving choices about their relative sizes are possible.  
We follow Pascalutsa and Phillips [63] and use the ratio of the two scales, $\delta$, as our small parameter. In Compton scattering this has the advantage that it allows us to differentiate between the kinematic regime where $\omega \sim m_\pi$ and that where $\omega \sim M_\Delta - M_{\rm N}$.
We obtain an amplitude which at low energies is complete
at $\calO(e^2\delta^4)$, where the leading (Thomson) amplitude is $\calO(e^2)$, while around the Delta
resonance the amplitude is complete at
order $\delta$ relative to leading order.

We determine $\alphae$ and $\betam$ by fitting to the proton
Compton-scattering database established in Ref.~\cite{Gr12}. To that end, we
constrain the parameters of the M1 $\gamma \mathrm{N} \Delta$ coupling by the
data around the Delta-resonance, and then extract polarisabilities from the
data at lower energy. We find that an excellent fit can be obtained by
including one contact operator which corresponds to a short-distance
contribution to the spin-polarisability $\gammamm$---even though, strictly speaking, this effect only appears at one order beyond that to which our calculation is complete. 
In doing this we are departing from the strict EFT expansion, but the same strategy is necessary at lower order where, once the Delta is included, $\alphae$ and $\betam$ must be promoted~\cite{Hi04}.

This article is structured as follows. First, we define the theoretical
ingredients of our calculation: the framework in
Sec.~\ref{sec:lagrangian}; the
power counting in the different kinematic regimes and how to arrive at an
amplitude which is valid across regimes in Sec.~\ref{sec:PCregimes}; explicit
results for the $\gammaN \Delta$ vertex corrections in
Sec.~\ref{sec:gNDvertex}; the prescription for the kinematically-correct
position of the pion-production threshold in Sec.~\ref{sec:thresholds}; and
constraints from sum rules and other processes in Sec.~\ref{sec:constraints}.
Sec.~\ref{sec:results} is devoted to the fit itself.  We present our results in
Sec.~\ref{sec:fits}, followed by a discussion of details, alternative
scenarios and convergence issues as well as, importantly, an estimate
of the residual theoretical uncertainties in Sec.~\ref{sec:details}.
After the Summary and Outlook of Sec.~\ref{sec:conclusion}, an Appendix gives references for the explicit forms of the components of the amplitudes.
Preliminary versions of our findings appeared in Refs.~\cite{McG09,Ph12,Gr12}.

\section{Theoretical Ingredients}
\label{sec:theory}

\subsection{Framework}
\label{sec:lagrangian}

The standard decomposition of the Compton-scattering amplitude  $T$  is given in 
Refs.~\cite{McG01,Gr12}, as are the expression relating $|T|^2$ to the differential cross sections 
 in  the centre-of-mass (cm) and laboratory (lab)
frames.  For our calculations we use the Breit  frame in which the photon does not transfer energy to the proton. See Refs.~\cite{Ba98,Gr12} for  details of kinematics in the various frames.

As detailed in Ref~\cite{Gr12}, the Compton amplitude can be split into Born and structure terms, the former 
arising (in a covariant framework) from nucleon- and pion-pole graphs, and the latter from all other effects.   
A low-energy expansion of the structure amplitude can be used to define  the electric and
magnetic scalar polarisabilities $\alphae$ and $\betam$, as well as the four spin polarisabilities.
In the Breit frame,
such an $\omega$-expansion gives, for the polarisability contributions to the structure amplitude,
\begin{align}
  \label{eq:Asinw}
    \frac 1 {4\pi}  T^{\textrm{pol}}&= (\alphaep + \betamp \, \cos\theta)\omega^2\,(\vec{\epsilon}\,'^*\cdot \vec{\epsilon})  -
  \betam \omega^2  \;(\vec{\epsilon}\,'^*\cdot\hat{\vec{k}})\;(\vec{\epsilon}
    \cdot\hat{\vec{k}}')
     \nonumber\\&
     -(\gammaeep+\gammaemp + (\gammamep+\gammammp) \, \cos\theta) \,\omega^3\,\ii \,\vec{\sigma}\cdot
    (\vec{\epsilon}\,'^*\times\vec{\epsilon}
    ) \nonumber\\&
    +
  (\gammammp-\gammamep)\,\omega^3\,  \ii \,\vec{\sigma}\cdot
    (\hat{\vec{k}}'\times\hat{\vec{k}}
    ) (\vec{\epsilon}\,'^*\cdot\vec{\epsilon}) 
    \nonumber\\&
     +\gammammp \,\omega^3\,\ii
    \,\vec{\sigma}\cdot
    [
    (\vec{\epsilon}\,'^*\times\hat{\vec{k}}
    )\,(\vec{\epsilon}\cdot\hat{\vec{k}}') -
    \vec{\epsilon}\times\hat{\vec{k}}'
    )\, (\vec{\epsilon}\,'^*\cdot\hat{\vec{k}})
    ]
    \nonumber\\&
    +\gammaemp \, \omega^3\,  \ii\,\vec{\sigma}\cdot
    [
    (\vec{\epsilon}\,'^*\times\hat{\vec{k}}'
    )\, (\vec{\epsilon}\cdot\hat{\vec{k}}') -
    (\vec{\epsilon} \times\hat{\vec{k}} 
    )\, (\vec{\epsilon}\,'^*\cdot\hat{\vec{k}})
    ] +\calO(\w^4),
    \end{align}
where $\vec{k}$ and $\vec{\epsilon}$ ($\vec{k}'$ and $\vec{\epsilon}\,'$)
are the three-momentum and polarisation vector of the incoming (outgoing)
photon, and $\hat{\vec{k}}$ ($\hat{\vec{k'}}$) unit vectors pointing in the
direction of $\vec{k}$ ($\vec{k}'$).

The  polarisabilities used in
Eq.~(\ref{eq:Asinw}) can be generalised to define dynamical,
i.e.~energy-dependent, polarisabilities~\cite{Griesshammer:2001uw,Hi04}. (These should not
be confused with the generalised polarisabilities that can be accessed in
virtual Compton scattering.)  This is equivalent to a multipole decomposition
of the structure (i.e.~non-Born) part of the Compton amplitude.  
For further discussion of the definition and usefulness of
energy-dependent/dynamical polarisabilities, see Ref.~\cite{Gr12}.  Here, 
however, we work with the full amplitudes rather than a truncated 
multipole decomposition.

We calculate the amplitudes within the framework of  \ChiEFT. Since the heavy-baryon $\piN$ Lagrangian is well-established, see
e.g.~Refs.~\cite{BKM, SS12}; we do not reproduce it here.
We use terms up to fourth order in  ${\cal L}_{\piN}$, using the conventions of Refs.~\cite{BKM,FMMS,MMS}.
The terms relevant to Compton
scattering at the order to which we work are given in Ref.~\cite{Gr12}.
Only the leading terms of ${\cal L}_{\pi \pi}$ are required, except for the anomalous part of
${\cal L}_{\pi \pi}^{(4)}$ which governs $\pi^0\to\gamma\gamma$.

One set of terms though are worth including, since we will refer to them again.
At $\calO(e^2 P^2)$, divergent loops need to be renormalised by operators that
occur in ${\cal L}_{\piN}^{(4)}$. Those relevant for Compton scattering are:
\begin{equation}
  {\cal L}_{\piN}^{(4),\text{CT}}=2 \pi e^2H^\dagger \Bigl[ 
  {\textstyle \frac{1}{2}}(\delta\betam^{(\mathrm{s})}
    +\delta\betam^{(\mathrm{v})}\tau_3)
  g_{\mu\nu}-\left[(\delta\alphae^{(\mathrm{s})} +
    \delta\betam^{(\mathrm{s})})+(\delta\alphae^{(\mathrm{v})} +
    \delta\betam^{(\mathrm{v})})\tau_3\right] v_\mu v_\nu\Bigr]F^{\mu\rho}
  F^{\nu}_{\; \; \rho}H . \label{eq:LpiN4} 
\end{equation}
They translate into additional, energy-independent contributions,
$\delta\alphae$ and $\delta\betam$, to the electric and magnetic polarisabilities
which are isoscalar or isovector.

For the $\Delta(1232)$ resonance, our calculation uses two variants.  As we
discuss in Secs.~\ref{sec:pc-delta} and~\ref{sec:gNDvertex}, we perform a
fully Lorentz-covariant calculation of the Delta-pole diagrams.  
$ {\cal L}_{\Delta}^{(1)}$ and $ {\cal L}_{\pi \mathrm{N}\Delta}^ {(1)}$ 
are taken from Pascalutsa and Phillips~\cite{PP03}.
However for loops we use corresponding terms from the heavy-baryon 
reduction of Hemmert et al.~\cite{He96} which are reproduced in Ref.~\cite{Gr12}.  The $\piN \Delta$ coupling constant in the latter is denoted  $\gpiNDelta$, which corresponds to $h_A/2$ in the former  (but see later for a discussion of the appropriate values to use in the two cases).

Because we will refer repeatedly to the $\gammaN \Delta$ couplings, though, we reproduce 
the relevant terms here.  Since we work to NLO in the
domain where the photon energy and the Delta-nucleon mass splitting
$\DeltaM\equiv \MDelta-\MN$ are comparable, $\omega \sim \DeltaM$, we need
selected terms from the Lagrangian up to third order~\cite{Ge99}:
\begin{align}
  {\cal L}_{\gamma \text{N}\Delta}^{\text{HB}, (2)}&=-\frac{\ii e
    b_1}{\MN}\Big(H^\dagger S_\rho F^{\mu\rho}\Delta^3_\mu -
  (\Delta^3_\mu)^\dagger S_\rho F^{\mu\rho}
  H\Big) \label{eq:subleading}\\
  {\cal L}_{\gamma \text{N} \Delta}^{\text{HB}, (3)}&=-\frac{e b_2}{2\MN^2}
  H^\dagger \Big(\stackrel{}{S} \cdot \stackrel{\leftarrow}{D} F^{\mu \rho}
  v_\rho + F^{\mu \rho} v_\rho S \cdot D\Big) \Delta_\mu^3 + \frac{e D_1}{4
    \MN^2} \left[H^\dagger v^\alpha S^\beta \langle \tau^3 [D_\mu,f^+_{\alpha
      \beta}] \rangle \Delta_\mu^3 + \rm{H.c.}\right],
  \label{eq:subsubleading}
\end{align}
 where the
field $\Delta_\mu^a$ of mass $\MDelta$ is an iso-quadruplet heavy-baryon
reduction of the Rarita-Schwinger field $\Psi_\nu^a$~\cite{He98},
 $a$ ($\mu$) being the index of the isovector (vector) coupled to the
isospinor (spinor).  $e=-|e|$ is the electron charge. All other symbols have their standard meaning. Terms generated by the heavy-baryon reduction are omitted in
Eq.~\eqref{eq:subsubleading}.
The convention used here is in accord with that of
Refs.~\cite{hhkk,Hildebrandtthesis,Hi04,Gr12}.  In
Eq.~\eqref{eq:subsubleading} we have identified Ref.~\cite{Ge99}'s $b_6 \equiv
-2 b_2$.

The $b_1$ interaction is of order $e
\omega$, while both the $b_2$ and $D_1$ terms give rise to couplings which are of
order $e \omega^2$. 

The corresponding form in the covariant version is~\cite{PP03}: 
\begin{align}
   {\cal L}_{\gammaN \Delta}^{\text{PP}, (2)}
  &=\frac{3e}{2\MN(\MN+\MDelta)}\Big[\bar\psi (\ii \gM
  \tilde F^{\mu\nu}-\gE \gamma_5
  F^{\mu\nu})\partial_\mu\Psi^3_\nu - \bar\Psi^3_\nu
  \overleftarrow{\partial}_\mu(\ii \gM \tilde F^{\mu\nu}\-\gE\gamma_5 F^{\mu \nu}) \psi
  \Big],
  \label{eq:PPEMLagrangian}
\end{align}
where
$\gamma^{\mu \nu}=\frac{1}{2}[\gamma^\mu,\gamma^\nu]$,
$\gamma^{\mu\nu\alpha}=\frac{1}{2} \{\gamma^{\mu \nu},\gamma^\alpha\}$,
$\gamma_5=\ii\gamma^0\gamma^1\gamma^2\gamma^3$, 
$\tilde F^{\mu\nu}=\frac 1 2 \epsilon^{\mu \nu \rho \sigma} F_{\rho \sigma} $ and $\epsilon_{0123}=+1$. 
 The (leading)
terms involving the electromagnetic field that emerge upon heavy-baryon
reduction of this Lagrangian are equivalent to those in
 Eqs.~\eqref{eq:subleading} and (\ref{eq:subsubleading}), provided we make the
identification
\begin{equation}\label{eq:gmge-b1b2}
  \gM= b_1(1+\MDelta/\MN)/3,\qquad \gE= b_2(1+\MDelta/\MN)/3.
\end{equation}
In this presentation, we use the relativistic variant of the
$\gammaN \Delta$ couplings, but quote results in terms of $b_1$ and
$b_2$, using this conversion formula throughout, since these are more widely
used.
The $s$- and $u$-channel Delta-pole diagrams then reproduce the results given
in the Appendices of Ref.~\cite{PP03}.
However, the Lagrangians of Refs.~\cite{PP03,PV06} do not include a structure
that reduces to that with coefficient $D_1$ in Eq.~(\ref{eq:subsubleading});
this point will play a role in Sec.~\ref{sec:gNDvertex}.

\subsection{Power counting regimes}
\label{sec:PCregimes}

As mentioned in the Introduction, three typical low-energy scales exist in
Compton scattering with a dynamical $\Delta$(1232): the pion mass
$\mpi\approx 140\;\MeV$ as the typical chiral scale; the Delta-nucleon mass
splitting $\DeltaM \equiv \text{Re}[\MDelta]-\MN\approx 290\;\MeV$; and the
photon energy $\omega$. Each provides a small, dimensionless expansion
parameter when it is measured in units of a natural ``high'' scale
$\Lambda\gg\DeltaM,\mpi,\omega$ at which \ChiEFT with explicit $\Delta(1232)$
degrees of freedom can be expected to break down because new degrees of
freedom enter.  Typical values of $\Lambda$ are the masses of the $\omega$ and
$\rho$ as the next-lightest exchange mesons (about $700\;\MeV$), so that
\begin{equation}
  P \equiv \frac{m_\pi}{\Lambda}\approx 0.2, \qquad 
  \epsilon \equiv \frac{\MDelta-\MN}{\Lambda}\approx0.4.
\end{equation}
While a threefold expansion of all interactions is possible, it is more
convenient to approximately identify some scales so that only one
dimensionless parameter is left.  In order to assign a unique index to each
graph, we therefore follow the $\delta$-expansion of Pascalutsa and
Phillips~\cite{PP03}. Since $m_\pi \ll \MDelta-\MN$, it takes
advantage of the numerical proximity $P\approx\epsilon^2$ in the real world to
define
\begin{equation}
  \delta\equiv
  \frac{\MDelta-\MN}{\Lambda}=\left(\frac{m_\pi}{\Lambda}\right)^{1/2}\qquad,
  \label{eq:delta}
\end{equation}
i.e.~numerically $\delta\approx\epsilon\approx P^\frac{1}{2}$.  The approach
then separates the calculation into two different regimes, depending on the
photon energy. In \textbf{regime I}, $\omega\lesssim\mpi\approx 140\;\MeV$ is
``low'', so that one counts $\omega\sim\mpi\sim\delta^2\Lambda$ like a
chiral scale. At higher energies (\textbf{regime II}, $\omega\sim\DeltaM\approx
300\;\MeV$), one counts $\omega\sim\DeltaM\sim\delta\Lambda$.

This power counting accommodates the fact that Compton scattering changes
qualitatively with increasing energy. In regime II, the photon carries enough
energy to excite a $\Delta(1232)$ intermediate state whose large width and
strong $\gammaN\Delta$ coupling makes it dominate the amplitudes. At
lower energies, the Delta should play a less pronounced role. This will be
confirmed in the list of all contributions in the following sub-sections.

\subsubsection{Regime I: $\omega \sim m_\pi$}
\label{sec:pc-mpi}
\begin{figure}[!t]
  \centerline{\includegraphics*
    {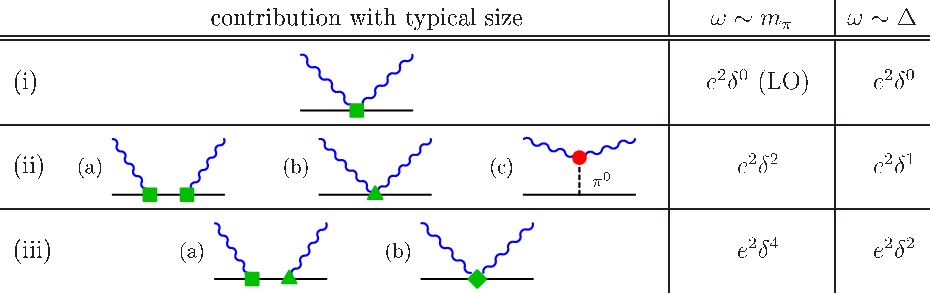}}
  \caption{\label{fig:protborn} (Colour online) Tree diagrams that contribute
    to Compton scattering in the $\epsilon\cdot v=0$
    gauge
    , ordered by the typical size of their contributions in the two regimes
    $\omega\sim\mpi\sim\delta^2$ and $\omega\sim\DeltaM\sim\delta$,
    respectively.  The leading-order contribution in a particular regime is
    indicated by (LO).  The vertices are from: ${\cal L}_{\piN}^{(1)}$ (no
    symbol), ${\cal L}_{\piN}^{(2)}$ (square), ${\cal L}_{\piN}^{(3)}$
    (triangle), ${\cal L}_{\piN}^{(4)}$ (diamond), ${\cal L}_{\pi\pi}^{(4)}$
    (disc). Permuted and crossed diagrams not shown.}
\end{figure}

\begin{figure}[tbp]
  \begin{center}
    \includegraphics
    {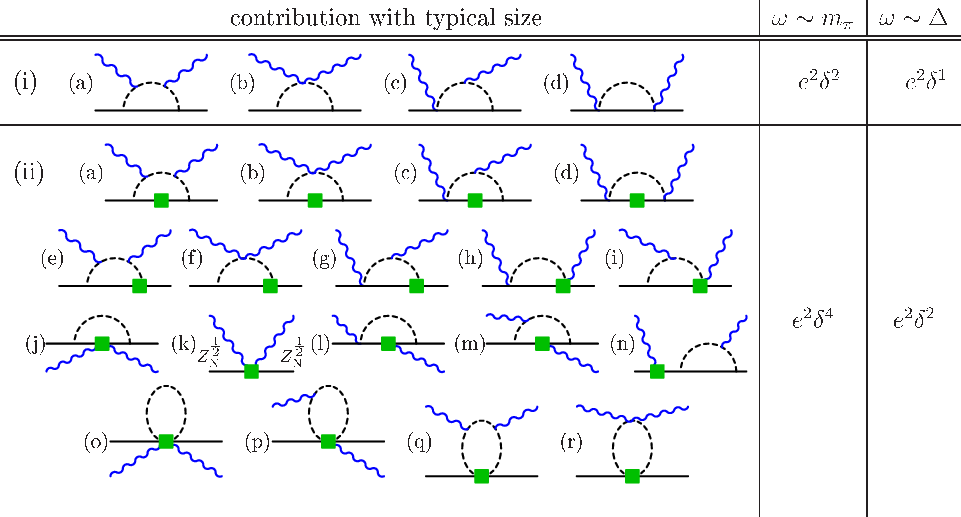}
  \end{center}
 \caption{\label{fig:protloop} (Colour online) Pion-nucleon loop diagrams.
Notation as in Fig.~\ref{fig:protborn}. Permuted and crossed
diagrams not shown.}
\end{figure}

\begin{figure}[tbp]
  \begin{center}
    \includegraphics
    {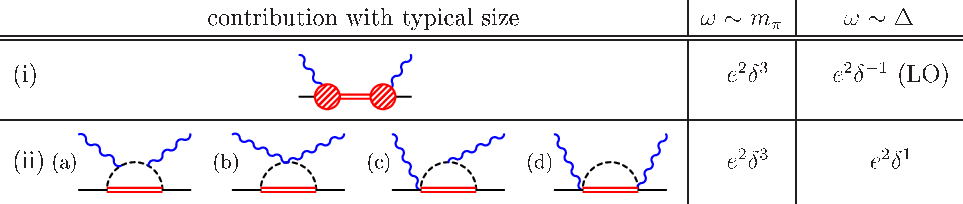}
  \end{center}
 \caption{\label{fig:protdelta} (Colour online) $\Delta(1232)$ and $\Delta \pi$ loop diagrams.
Notation as in Fig.~\ref{fig:protborn}, also double line:
$\Delta(1232)$; shaded blob: $\piN\Delta$ couplings including vertex
corrections, as detailed in Sect.~\ref{sec:gNDvertex} and
Fig.~\ref{fig:protdelta} below. The given order for (i) of $\calO(e^2\delta^{-1})$  for $\w\sim\DeltaM$ applies only to the leading coupling at the vertices; other contributions will contribute at higher order. Permuted and crossed diagrams not shown.
However, the $s$- and $u$-channel Delta-pole graphs in (i) occur at different
orders in regime II, as detailed in the text.}
\end{figure}

We first discuss
the graphs  that contribute up to the order to which we work, Figs.~\ref{fig:protborn} to~\ref{fig:protdelta},  for photon energies $\omega\sim\mpi$. References for the actual form of the resulting amplitudes are given in Appendix \ref{sec:appendix}.

The first class consists of the ``tree'' graphs of
Fig.~\ref{fig:protborn}, For contributions
without an explicit $\Delta(1232)$, the only expansion scale in this regime is
$P\equiv\delta^2$, and the power counting is that of HB$\chi$PT, with only
even powers of $\delta$ contributing~\cite{PP03}.
Since we use the gauge $v\cdot A=0$,  direct $\gammaN$ couplings do not occur at lowest order.  
Thus
in regime I, the leading, $\calO(e^2\delta^0\sim e^2P^0)$, contribution is the ``seagull'' diagram Fig.~\ref{fig:protborn}(i) which gives the Thomson amplitude; the rest of the diagrams give the pion-pole contribution, and the nucleon Born amplitude expanded to $\calO(1/\MN^3)$.   The 4th-order seagull graph, diagram
(iii)(b), however, also contributes to the structure amplitude in the form of the  short-distance contributions $\delta\alphae,\;\delta\betam$ to the scalar
polarisabilities of Eq.~\eqref{eq:LpiN4}.  

The second class of contributions comprises the pion-loop diagrams
of Fig.~\ref{fig:protloop}.  The
leading contributions  are  $\calO(e^2\delta^2\sim e^2P)$ and give the polarisability contributions of Eq.~\eqref{eq:BKM}.
In dimensional regularisation the full amplitude  contains only two divergences at order
$\calO(e^2\delta^4)$, which are cancelled by  $\delta\alphae$ and $\delta\betam$.  
The finite parts of these LECs encode contributions to the
polarisabilities from mechanisms other than soft-pion loops or Delta
contributions, i.e.~from short-distance effects. In contrast, the spin
polarisabilities are still parameter-free predictions at this order. Two- and higher-loop diagrams and contributions from higher-order
Lagrangians enter only at $\calO(e^2\delta^6)$, and are
thus suppressed.
(In Sec.~\ref{sec:thresholds} we discuss the prescription which  puts the pion-production threshold at the kinematically correct position \cite{McG01,Hi04}.)

The diagrams in the third class contain a dynamical $\Delta(1232)$, as listed
in Fig.~\ref{fig:protdelta}.
The $\Delta(1232)$ propagator in these graphs is 
\begin{equation}
  \label{eq:deltaprop-regimeI}
  S^{(0)}_\Delta(\omega_\Delta\sim\mpi)\propto\frac{1}{\DeltaM\pm\omega_\Delta},
\end{equation}  
where $\omega_\Delta$ is the kinetic energy of the Delta line, which is dominantly $\calO(\mpi)$ in  regime I. Since $\omega_\Delta\ll\DeltaM\sim\delta$, the $\Delta(1232)$ propagator
scales as $\delta^{-1}$, in contrast to the nucleon propagator, $\ii /\omega_\p$, which scales as
$P^{-1}\sim\delta^{-2}$. Diagrams with a Delta  are thus suppressed by one
order in $\delta$ relative to the corresponding nucleon diagrams, and hence start at $\calO(e^2\delta^3)$.
Here, we made the argument using the heavy-baryon propagator for the Delta, as
it is more transparent, even though we actually choose to compute the pole
graphs using a relativistic propagator, as described in
Sec.~\ref{sec:pc-delta}.  The $\pi \Delta$ loops {\em are} computed using the heavy-baryon propagator, and this ostensibly generates an inconsistency. However, the difference between the results with heavy-baryon and relativistic propagators  is higher order in the expansion in $1/\MN\sim1/\Lambda$, and is numerically small~\cite{Le12}, so the inconsistency thereby introduced is insignificant.

Corrections to these Delta-contributions from ${\cal L}_{\piN \Delta}^{(2)}$
add a power of $P\sim\delta^2$ and thus enter only at $\calO(e^2\delta^5)$. There
are therefore no corrections to the amplitude from $\Delta(1232)$ effects at
$\calO(e^2 \delta^4)$ in this kinematic regime.

\subsubsection{Regime II: $\omega \sim \MDelta - \MN$}
\label{sec:pc-delta}

\begin{figure}[tbp]
  \centerline{\includegraphics
    {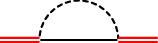}}
  \caption{(Colour online) The leading-order contribution to the
    $\Delta(1232)$ self-energy $\Sigma_\Delta$ in regime II ($\omega \sim \DeltaM$).}
  \label{fig:deltaresummation}
\end{figure}

In this regime, the $\Delta(1232)$ propagates close to its mass-shell, and its
non-zero width is predominantly generated by $\piN$ loops. The details of the
$\delta$ counting for this case were derived in Ref.~\cite{PP03} and are
briefly repeated here. The leading effect comes from re-summing the leading
one-Delta-reducible diagrams via a Dyson equation to obtain a dressed Delta
propagator, so that Eq.~(\ref{eq:deltaprop-regimeI}) becomes
\begin{equation}
  \label{eq:deltaprop-regimeII}
  S^{(0)}_\Delta(\omega\sim\DeltaM)\propto\frac{1}{\DeltaM + \ii \,\Im\Sigma_\Delta-\omega}.
\end{equation}  
 Since the kinetic energy flowing through the $\Delta(1232)$
propagator in this region is $E\sim\delta$ and the $\piN\Delta$
vertex  scales as $\delta$, the leading self-energy
contribution, Fig.~\ref{fig:deltaresummation}, scales as $\delta^3$.
Written as a function of Mandelstam $s$, its imaginary part is:
\begin{equation}
  \label{eq:deltaself}
  {\rm Im}[\Sigma_{\Delta}(s)]=-\left(\frac{\gpiNDelta}{2 \MDelta}\right)^2 \frac{(\sqrt{s} + \MN)^2 - m_\pi^2}{48 \pi \MDelta^2}\; k^3\, \theta(k),
\end{equation}
where $\theta(k)$ is the Heaviside function and $k$ the centre-of-mass momentum
of the pion in the $\piN$ system:
\begin{equation}
 k=\sqrt{\frac{[s-(\MN+m_\pi)^2][s-(\MN-m_\pi)^2]}{4s}}.
\end{equation}
Following Ref.~\cite{PP03}, we keep only the imaginary part of $\Sigma_\Delta$
and absorb the real part into $\MDelta$ and into the Delta wave-function
renormalisation. Note that we quote the relativistic result for the Delta self energy, for reasons to be explained.

After it is dressed by the self-energy (\ref{eq:delta}), the
Delta propagator \eqref{eq:deltaprop-regimeII} scales as $\delta^{-3}$.  This
is different from the scaling of the propagator 
\eqref{eq:deltaprop-regimeI} for $\omega\ll\DeltaM$ because as $\omega$
becomes comparable to $\DeltaM$, there is a region where the difference
$\omega-\DeltaM$ scales as $|\omega - \DeltaM| \sim \Lambda \delta^3$. With
$\Lambda\approx 700\;\MeV$, this is numerically the case for
$|\omega-\DeltaM|\approx 50\;\MeV$. For photon energies this close to the
$\Delta(1232)$-nucleon mass-splitting, all terms in the Dyson series are of
the same order, $\Sigma_\Delta\sim\delta^3$, so that the resummation of the
Delta width is mandated by the power counting. This conforms with the
reasoning that regime II should be limited to the region about the resonance
position, with its width approximately given by the width of the resonance.

The fact that, relative to regime I, the photon energy is enhanced by one order
to $\omega\sim\delta$, and the Delta propagator near resonance is enhanced by two
orders, changes the relative importance of the contributions to Compton
scattering in regime II. The leading order is $\calO(e^2 \delta^{-1})$, and
consists of the
$s$-channel tree-graph, Fig.~\ref{fig:protdelta}(i), which contains the
leading M1 coupling $b_1$ and the dressed $\Delta(1232)$ propagator~\eqref{eq:deltaself}.

Since we already use the relativistic self energy, NLO ($\calO(e^2
\delta^{0})$) contributions to the Compton amplitude come only from Thomson ``seagull" of  Fig.~\ref{fig:protborn}(a), and from the
$\gammaN  \Delta$ vertex corrections, symbolically included in the
shaded blob of Fig.~\ref{fig:protdelta}(i). These appear as both loop and
counterterm effects, including one $E2$ coupling $b_2$, and will be discussed
in Sec.~\ref{sec:gNDvertex}.

In the u-channel pole and $\pi \Delta$-loop graphs the Delta propagator is not enhanced, but counts as $\delta^{-1}$, as in regime I. These, along with the $\piN$ loops, all start at $e^2\delta$ (\NXLO{2}) in regime II. Higher-order couplings, together with two-loop contributions, also contribute at this order.
 
So far our discussion of power counting has been based on HB$\chi$PT.  
A strict application would give the peak of the resonance at $\w=\DeltaM$, with $\w$ the Breit-frame photon energy.  The correct position, $\sqrt{s}-M_N=\DeltaM$, would then be restored perturbatively as an expansion in $1/\MN$.  The cross section in this regime is dominated by the Delta resonance, and is very sensitive to its exact location. At a minimum, those corrections which ensure the correct 
real part of the resonance energy must be resumed. Doing this means that we avoid purely kinematic---and hence artificial---shifts of the polarisabilities from one order to the next. (See Sec.~\ref{sec:thresholds} for the analogous procedure for the pion loops.) Note that, even with the correct real part of the resonance energy ensured, the full EFT amplitude up to NLO in the resonance region includes a number of $1/\MN$-suppressed vertex effects. 
Furthermore, despite the fact that $\sqrt{s}-\MN \ll \MN$ at the resonance, the strong dependence of $\Sigma_\Delta(s)$ on $k$ (see Eq.~\eqref{eq:deltaself}) means that the
relativistic and non-relativistic widths differ by a factor of 2 (for a fixed value of $\gpiNDelta$). 

An efficient way of dealing with these issues in practice is to use a covariant expression for the Delta-pole graphs, based on the Lagrangian \cite{PP03}. This automatically takes care of both recoil and vertex corrections, by including effects which are higher-order in the power counting than the order to which we are working; except for the correct kinematics this cannot improve our agreement with data, but---as long as the power counting is valid---it should not hinder it either.  In the same spirit, we also determine the resonance
  parameters $\MDelta$ and $\gpiNDelta$ from the Breit-Wigner peak and width,
  the latter via the relativistic formula. 

Since the kinematic shifts are irrelevant at these energies in the $\pi\Delta$ loops, we retain the heavy-baryon formulation for those~\cite{He97,hhkk}.
 
\subsubsection{Unified description of regimes I and II}
\label{sec:unifiedpc}

While separate sets of amplitudes for regimes I and II are possible, matching
them smoothly in the transition region would be cumbersome. It is more
convenient to work with one set of amplitudes which is applicable in both
regimes simultaneously. We therefore choose to include \emph{all} diagrams
which contribute to \NXLO{4} ($\calO(e^2\delta^4)$) in regime I or to NLO
($\calO(e^2\delta^0)$) in regime II, even if they contribute at higher orders
in the other regime. This is indeed the set of diagrams discussed above, Figs.~\ref{fig:protborn} to \ref{fig:protdelta}, with a
relativistically resummed Delta propagator for the tree amplitudes of
Fig.~\ref{fig:protdelta}(i).
Moreover,
each set is gauge and renormalisation-group invariant by
itself.  Therefore, adding them does not lead to conceptual problems, but
neither does it improve the accuracy of the 
calculation.  It is merely convenient to describe Compton scattering on the
proton with one set of amplitudes from zero energy up to the Delta
resonance.  The accuracy in each regime is still given by the highest order
at which all contributions are included.  In regime I ($\omega\lesssim\mpi$),
that order is $\calO(e^2 \delta^{4})$ or 4 orders beyond LO, and the relative
systematic uncertainty of the Compton cross section can thus \emph{a priori}
be assessed as $\delta^{5}\approx[\frac{2}{5}\dots\frac{1}{2}]^{5}\approx1$ to
$3\%$. In regime
II ($\omega\sim\DeltaM$), that is $\calO(e^2 \delta^{0})$, i.e.~NLO, with a
relative \emph{a priori} error of
$\delta^{2}\approx[\frac{2}{5}\dots\frac{1}{2}]^{2}\approx15$ to $25\%$.

Since the amplitudes are therefore less accurate at high energies, we choose
in the following to analyse in detail the data lying in regime I, and then
check the consistency 
with experiments in the Delta-resonance region.

We finally stress that the regimes of applicability of the different counting
schemes are estimates: transition from one regime to the next is not abrupt
but gradual, and the above considerations provide only typical \emph{a priori}
error estimates. Convergence of the \ChiEFT expansion order-by-order must be
checked carefully so as to reliably assess theoretical uncertainties.
Ultimately, comparison to data will help to decide whether the theory
accurately describes Compton scattering.

\subsection{$\gammaN  \Delta$ vertex corrections}
\label{sec:gNDvertex}

As alluded to in Sec.~\ref{sec:pc-delta}, corrections which contribute to the
vertices of the Delta-pole graph of Fig.~\ref{fig:protdelta}(i) are, in the
low-energy regime, of higher order than that to which we work.  However in the
medium-energy region they are promoted to start at NLO; see
Fig.~\ref{fig:vertexstructure}:
\begin{figure}[tbp]
  \centerline{\includegraphics[width=.6\linewidth]{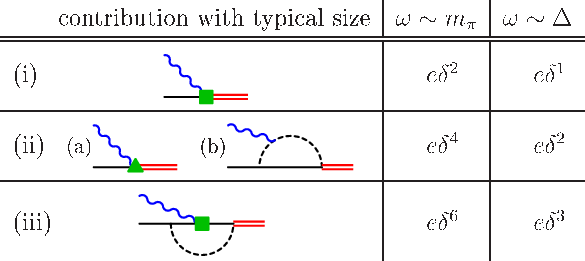}}
  \caption{(Colour online) Contributions to the $\gammaN\Delta$ coupling of Fig.~\ref{fig:protdelta}, ordered by their typical size 
    for $\omega\sim\mpi\sim\delta^2$ and $\omega\sim\DeltaM\sim\delta$,
    respectively.  Notation as in Figs.~\ref{fig:protborn}
    and~\ref{fig:protdelta}; $\gammaN \Delta$ vertices from ${\cal
      L}_{\gammaN \Delta}^\text{PP}$, Eq.~\eqref{eq:PPEMLagrangian},
    at $\calO(e\delta^2)$ proportional to $\gM$ (i.e.~$b_1$; square), and at
    $\calO(e\delta^3)$ proportional to $\gE$ (i.e.~$b_2$; triangle). For
    diagrams (ii)(b) and (iii), the order at $\omega\sim\DeltaM$ refers to the
    imaginary part only; the real part is demoted. Permuted and crossed
    diagrams not shown.}
  \label{fig:vertexstructure}
\end{figure}
the vertex corrections are $\calO(e \omega^2)$, whereas the dominant ($b_1$)
vertex is $\calO(e \omega)$. Thus pion-loop effects in the
$\gammaN \Delta$ vertex contribute to the $\calO(e^2 \delta^0)$ part
of the Delta-pole graph.  They are therefore predicted to be more important in
the vicinity of the resonance than even the basic $\piN$-loop graphs depicted
in Fig.~\ref{fig:protloop}. Since we use $\chi$EFT to fit Compton data in the
resonance region---especially to constrain the M1 $\gammaN \Delta$
coupling---we include $\piN$ vertex loops in our calculation in order to be
complete to NLO in this region.  They have not been considered in recent
Compton-scattering calculations~\cite{LP09,LP10,Gr12}, although the importance of
the imaginary part of Fig.~\ref{fig:vertexstructure}(ii)(b) was recognised in the
original $\delta$-expansion study of Ref.~\cite{PP03}.

\begin{figure}[tbp]
  \centerline{\includegraphics
    {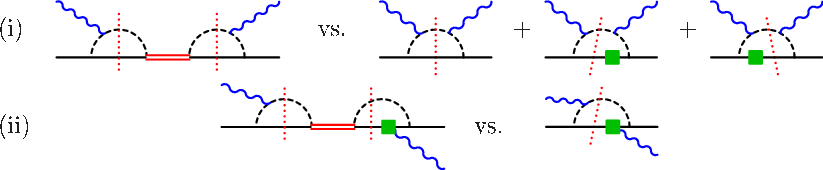}}
  \caption{(Colour online) Pairs of diagrams whose imaginary parts (with cuts
    indicated by dotted lines) are related by Watson's theorem at the Delta
    resonance.  Notation as in Figs.~\ref{fig:protborn}
    and~\ref{fig:vertexstructure}.}
  \label{fig:watson}
\end{figure}

Inclusion of these graphs is also necessary to satisfy Watson's
theorem~\cite{Wa54}, which requires that the pion photoproduction amplitudes
at the peak of the resonance are purely imaginary.  This implies that the real
contribution from $\gamma\pi$N Born diagrams must be cancelled by Delta-pole
graphs with a purely imaginary $\gammaN \Delta$ vertex that results
from $\piN$ loops. When these elements are assembled into Compton diagrams the
net effect is that, at resonance, there are cancellations between the
isospin-3/2 pieces of of the $\piN$ loop graphs of Fig.~\ref{fig:protloop} and
Delta-pole graphs with imaginary $\gammaN \Delta$ vertices from the
loops of Fig.~\ref{fig:vertexstructure}(ii) and (iii).  Diagrams which are
connected in this way are shown in Fig.~\ref{fig:watson}.

Since we are using heavy baryons for the calculation of the loops of
Fig.~\ref{fig:protloop}, but the covariant version for the Delta-pole graph,
we have a choice about whether or not to use the heavy-baryon expansion in
computing these vertices; neither choice is fully consistent, although the
inconsistencies are associated with higher-order effects in $\delta$-counting.
Thus we use the covariant formulation to accord with other aspects
of our treatment of the Delta, such as the $\piN$-loop which gives the width
(see Sec.~\ref{sec:pc-delta}).

Pascalutsa and Vanderhaeghen~\cite{PV06} computed both $\piN$-loop graphs of
 Fig.~\ref{fig:vertexstructure} at leading order in the covariant
formulation of baryon $\chi$EFT. Since their calculation was reported at the
$\Delta(1232)$ peak as a function of $Q^2$, while we are interested in the
vertices for real photons away from the $\Delta(1232)$ peak, for completeness
we give results for this case here.  (Note that in Eq.~(47) of
Ref.~\cite{PV06} $\alpha_\gamma$ is a misprint for $\beta_\gamma$.)

Both diagrams contribute to both the
  structures whose coefficients are $b_1$ and $b_2$ at tree level.  These
  corrections are labelled as $b_i^{(\pi)}$ ($i=1,2$) for the first diagram,
  Fig.~\ref{fig:vertexstructure}(ii)(b), since it contains the photon coupling
  to the pion; and as $b_i^{(\mathrm{N})}$ ($i=1,2$) for the second diagram,
  Fig.~\ref{fig:vertexstructure}(iii), since in it the photon couples to the
  nucleon:
\begin{align}
  b_1^{(\pi)} &= - \CDeltaN \! \int_0^1 \! \dd y \, y
  \!\int_0^{1-y} \!\!\!  \!\!\!\dd x\, \ln \left[\frac{{\cal M}_a^2}{\MN^2}\right] ,
  \nonumber \\
  b_2^{(\pi)} &= -b_1^{(\pi)} - 2\CDeltaN\! \int_0^1 \! \dd y\, y
  \!\int_0^{1-y}\!\!\!\!\!\!\dd x\, x \, [ (1 - y)\MN + (1 - x - y)W ]W\,
  {\cal M}_a^{-2},
  \nonumber\\
  b_1^{(\mathrm{N})} &= - \CDeltaN \!\int_0^1\!  \dd y \, y
  \!\int_0^{1-y} \!\!\!\!\!\! \dd x\, \left\{ 2\ln \left[\frac{{\cal
        M}_b^2}{\MN^2}\right] - \frac{ x y \MN ^2+ (1 - x)W \MN + x (1- x - y)
      W^2}{{\cal M}_b^{2}}\right\},
  \nonumber\\
  b_2^{(\mathrm{N})} &= -b_1^{(\mathrm{N})} + 2 \CDeltaN \!
  \int_0^1 \!\dd y \, y \!  \int_0^{1-y} \!\!\!\!\!\! \dd x \, \left[ x (1 - x
    - y) W + (1 - x - x y) \MN \right]W\,{\cal M}_b^{-2} ,
  \label{eq:vertex}
\end{align}
where $W=\sqrt{s}$, $\CDeltaN=2\ga\gpiNDelta \MN^3/((4\pi\fpi)^2\MDelta)$, and

\begin{align}
  {\cal M}_a^2 &= x(1- y)\MN ^2+ (1 - x)\mpi^2- x (1 - x - y) W^2-
  \ii\epsilon, \nonumber\\
  {\cal M}_b^2 &= (1 - x - x y)\MN^2 + x \mpi^2- x (1 - x - y)W^2-
  \ii\epsilon\end{align}
The imaginary parts of both diagrams can be calculated in closed form
\cite{PV06}; we have also checked them against Watson's theorem for
photoproduction.  In the heavy-baryon limit, the expressions for graph (ii)(b)
(in which the photon couples to the pion in the $\piN$ loop) agree with those
obtained by Gellas et al.\ \cite{Ge99}, and for (iii) (in which the coupling
is to the nucleon) with our own heavy-baryon calculation---with the
following proviso. In the covariant theory graphs (ii)(b) and (iii) are
treated as being of the same (leading) order.  However in the heavy baryon
formulation (iii) is higher order, suppressed by $1/\MN$, as is most easily
seen in the gauge $\epsilon\cdot v=0$ in which the leading photon-nucleon
coupling is the magnetic one.  At the same order the anomalous magnetic moment
enters, so that the $\gammaN$ coupling is proportional to the total magnetic
moment.  Since we work to the order at which the anomalous magnetic moment
enters for the direct pion loops, we do the same for the vertex loops, and
multiply $b_1^{(\mathrm{N})}$ by the isovector Dirac magnetic moment
$\muv=1+\kappav$.  Furthermore, even at this order in the heavy-baryon theory,
there is no contribution from graph (iii) to the electric coupling $b_2$, the
leading term in an expansion of $b_2^{(\mathrm{N})}$ from
Eq.~\eqref{eq:vertex} being $\calO(1/\MN^2)$.

Following Ref.~\cite{PV06} we subtract from the above results the real parts
at $W=\MDelta$,
so that the renormalised Lagrangian constants $b_1$ and $b_2$ are defined to
give the real part of the dressed $\gammaN \Delta$ couplings at the
$\Delta(1232)$ pole.  When we discuss fitting $b_1$ we mean
this quantity.

However, even after such a subtraction, both contributions to ${b}_1(W)$
contain $\log(\MN)$ dependences.  This corresponds to a residual logarithmic
dependence on the renormalisation-scale in the heavy-baryon case~\cite{Ge99}.
That could be absorbed (for (ii)(b)) by the HB Lagrangian LEC $D_1$ of
Eq.~(\ref{eq:subsubleading}), but no corresponding LEC was included in the
relativistic Lagrangian of Ref.~\cite{PV06}. Since the expressions are finite,
and we have no information on the LEC $D_1$, we do not consider the additional
effect from it.

In fact, in the vicinity of the resonance, the variation with $W$ of the real
part of each loops graph is formally of higher order than the imaginary part
(the argument parallels that of Ref.~\cite{PP03} when discussing the
self-energy of the Delta).
Moreover, the $b_i$'s also receive (real) contributions from diagrams with the
same topology as diagram (ii)(b), in which the baryon propagating in the loop
is a $\Delta(1232)$. A full computation of the vertex functions to
this 
order is however beyond the scope of this paper. And indeed, at NNLO in this
kinematic domain the Delta propagator also receives substantial two-loop
corrections~\cite{LvK09}. Therefore in what follows we discard the real part
of diagram (iii), even when, working to fourth order, we retain the imaginary
part for consistency with Watson's theorem.

To summarise, therefore, the dressed vertices we use are
\begin{align}
  b_i^{(3)}(W)&=b_i + \ii \;\Im[b_i^{(\pi)}(W)] +
  \Re[b_i^{(\pi)}(W)-b_i^{(\pi)}(\MDelta)] \nonumber\\\
  b_1^{(4)}(W)&=b_i^{(3)}(W)+\ii\;\muv \Im[b_1^{(\mathrm{N})}(W)],\qquad
  b_2^{(4)}(W)=b_2^{(3)}(W),
  \label{eq:bi}
\end{align}
where superscripts (3) and (4) indicate use in calculations accurateÊÊin the low-energy regionÊto
$\calO(e^2\delta^3)$ and $\calO(e^2\delta^4)$ respectively.

\begin{figure}[tbp]
  \centerline{\includegraphics[width=\linewidth]{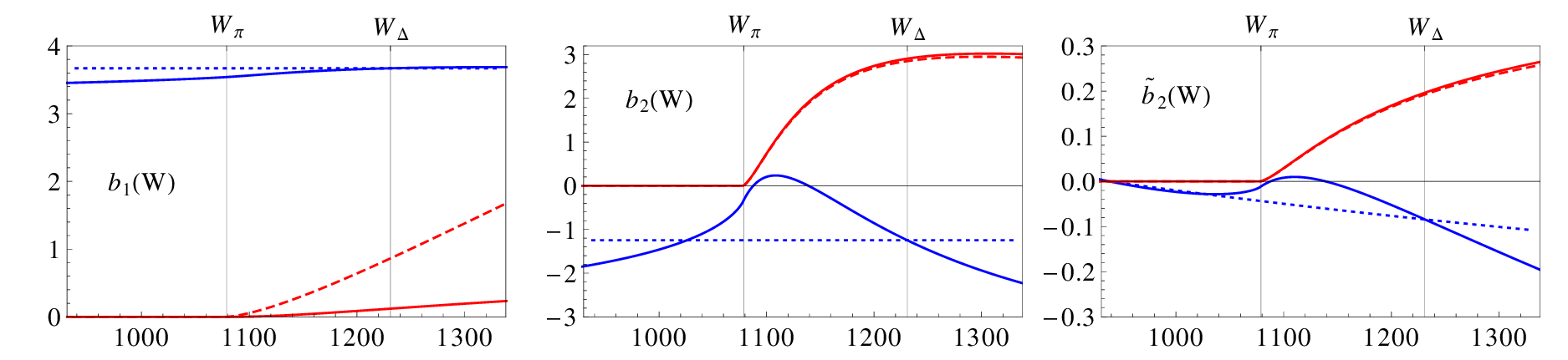}}
  \caption{(Colour online) The effects of including loop corrections to the
    $\gammaN \Delta$ vertex; in all graphs blue is the real part and
    red is the imaginary part (which vanishes below $W=\MN+\mpi$).  Dotted
    lines show no loop effects, solid the (renormalised) contribution of
    diagram (ii)(b) and dashed, both (ii)(b) and (iii) (imaginary parts only, with the
    $\muv$ factor for $b_1$). }
  \label{fig:formfactor}
\end{figure}

The results of this calculation are shown in Fig.~\ref{fig:formfactor}.  The
first two panels depict $b_1$ and $b_2$, with the dotted line giving the
constant (tree-level) result for comparison. When the effect of diagram (ii)(b) is
added the solid line is obtained (so these are $ b_i^{(3)}(W)$), and the
dashed line is the result when the imaginary part of graph (iii) is included as
well (with the $\muv$ factor for $b_1$, giving $ b_1^{(4)}(W)$). We see that
the hierarchy of effects predicted by the power counting (after
renormalisation) indeed holds:
\begin{equation}
  b_1 > b_1^{(\pi)},
\qquad
  b_2 \sim b_2^{(\pi)} \gg b_2^{(\mathrm{N})}
\end{equation}
(The inverted hierarchy for the loop contributions to $b_1$ has a double
origin: the contribution of (ii)(b) happens to be small, while the large value of
$\muv$ makes the contribution from (iii) unnaturally large.) As for
the relative size of $b_1$ and $b_2$, the ratio of multipoles E2/M1 is
\begin{equation}
  \frac{\rm E2}{\rm M1}=
  \frac{\tilde b_2(W)}{b_1(W) +\tilde b_2(W)},
\end{equation}  
where
  \begin{equation}
  \tilde b_2(W)=\frac{(W-\MN) b_2(W)}{2 (W + \MN)}, 
  \label{eq:E2overM1}
\end{equation}
since the operator multiplying $b_2$ contains an extra time derivative. The
quantity $\tilde b_2(W)$ is displayed in the third panel, where the different
scale on the $y$-axis makes it clear that the effect of $b_2$ on observables
is higher order than the constant leading term in $b_1$.  It was from the
ratio (\ref{eq:E2overM1}), evaluated at $W=\MDelta$, that Ref.~\cite{PV06}
obtained the result $b_2/b_1=-0.34$. We also use that result to fix the ratio
$b_2/b_1$.

We emphasise that the results we obtained for the variation in the real part
of $b_i(W)$ with $W$ are dependent on our particular definition of Delta-pole
piece. In particular, a redefinition of the $\Delta(1232)$ field would not
change the value of these functions at $W=\MDelta$, but could affect their
variation as we move away from the pole. Therefore the results shown in
Fig.~\ref{fig:formfactor} only have meaning within the context of the specific
${\cal L}_{\gammaN  \Delta}$ and ${\cal L}_{\piN \Delta}$ that we have
adopted here.

\subsection{Re-summing recoil corrections}
\label{sec:thresholds}

In the heavy-baryon version of one-nucleon \ChiEFT, interactions and
amplitudes are expanded in powers of $1/\MN$. As with the correct position of the
the $\Delta(1232)$ pole discussed at the end of Sec.~\ref{sec:pc-delta}, that
of the one-pion production threshold is thus only reached as a perturbative
expansion with higher-order terms. This has consequences for Compton
scattering since the opening of the photoproduction channel induces pronounced
cusps in the amplitudes, as discussed in detail in Ref.~\cite{Hi04}.
Pion production starts in Compton-scattering experiments at photon energy
$m_\pi + \frac{m_\pi^2}{2 \MN} \approx 150$ MeV in the lab frame. In contrast,
the HB amplitude neglects recoil by assuming a nucleon which is at LO
infinitely heavy, and thus puts the threshold at photon energy
$\mpi\approx140\;\MeV$.  The $\calO(e^2 P)$ HB$\chi$PT amplitude has a cusp at
this point, and that at $\calO(e^2 P^2)$ is singular there.  The consequences
of the cusp are seen even for $\omega$ below $m_\pi$, so getting it in the
right place is mandatory for an accurate description of data around the pion
threshold.

Various schemes exist in the literature to correct for these purely kinematic
effects by a resummation of higher-order terms that is motivated by a modified
power-counting in the vicinity of a threshold (see Ref.~\cite{Be93} for an
early discussion). All schemes agree within the order to which the \ChiEFT
calculation is performed, with each effectively trying to expand the
amplitudes not about $\omega=0$ but about the kinematically correct threshold
position. As in Ref.~\cite{Hi04} we replace the Breit- frame photon
energy, $\omega$, with the quantity $\omega_s$, defined via:
\begin{equation}
  \label{eq:threshold}
  \omega_s \equiv \sqrt{s}-\MN.
\end{equation}
This has the advantage that it implies changing the amplitude for $\omega\to0$
only starting at $\omega^2$. (Note that a slightly different substitution was adopted in Refs.~\cite{McG01,Be02,Be04}).

To preserve crossing symmetry we perform this substitution in all $s$- and $u$-channel (i.e.~crossed)
diagrams containing $\piN$ loops, Fig.~\ref{fig:protloop}. At $\calO(e^2 \delta^3)$ this is all
that is done. 
However at $\calO(e^2 \delta^4)$ we need to ensure a finite amplitude as we cross the
$\piN$ threshold. This requires that we do not double count the $\calO(e^2
P^2)$ effects which were already included in the $\calO(e^2 P)$ result via the
shift (\ref{eq:threshold}). Thus, if the straightforwardly computed HB$\chi$PT
loop amplitude is (in an obvious notation) $T^{(3)}(\omega) +
T^{(4)}(\omega)$, at $\calO(e^2 \delta^4)$ we compute:
\begin{equation}
  T^{(3)}\left(\omega_s \right) + T^{(4)}\left( \omega_s \right)
  - (\omega_s - \omega)^{(1)} \;\frac{\partial T^{(3)}(\omega_s)}{\partial
    \omega_s}, 
\end{equation}
where the superscript $(1)$ indicates that only the $\calO(\omega^2/\MN)$
piece of the frame-dependent factor $\omega_s - \omega$ is retained.  This
yields a finite result for the amplitude as it crosses the cusp, and that cusp
is now in the correct place regardless of the frame in which the Compton
amplitude is computed~\cite{Be93,McG01}.

We note again that---as long as we are not close to the specific locations of
analytic structure in the complex plane, namely the Delta pole and the $\piN$
cut---neither the resummation of recoil corrections to place the $\piN$
threshold in the right place, nor our use of a fully-covariant amplitude for
the Delta-pole graphs, add any new dynamics. In the heavy-baryon expansion,
all these kinematic effects reduce in either regime to corrections which are
higher than the orders with which we are concerned.

\subsection{Constraints from other processes and sum rules}
\label{sec:constraints}

Strictly speaking, there are two parameters at $\calO(e^2\delta^4)$ which can
only be determined from proton Compton scattering: the finite part of the
short-distance contributions $\delta\alphae$, $\delta\betam$ to the proton's
scalar electric and magnetic dipole polarisabilities. The
$\gammaN \Delta$ couplings $b_1$ and $b_2$ can in principle be
determined from the width and $E2/M1$ ratio of the $\Delta(1232)$ radiative
decay.

It is, however, useful to check the consistency of the values found for these
parameters against other constraints. For example, the Baldin sum rule relates
the sum of the scalar dipole polarisabilities at zero energy to an integral of
the weighted total dissociation cross section $\sigma_{\gammaN \to X}$
starting at the production threshold $\omega_\text{thr}$ via a dispersion
relation. We use the value from the recent global 
fit of~\cite{OdeL}:
\begin{equation}
  \label{eq:baldin}
  \alphaep + \betamp=\frac{1}{2\pi^2}\int\limits_{\omega_\text{thr}}^\infty
  \dd\omega\,
  \frac{\sigma_{\gammaN \to X}}{\omega^2}= 13.8\pm0.4,
\end{equation}
where the canonical units, $10^{-4}\;\fm^3$, for scalar polarisabilities, are to be understood here
and below.
The ``recommended value'' of Ref.~\cite{Sc05},
$13.9\pm0.3
$, which is based on an analysis of total photo-absorption cross sections,
differs from the value (\ref{eq:baldin}) by less than the uncertainty in the
sum-rule evaluation.

We also note that a recent reevaluation of the dispersion integral for
$\gammazerop$~\cite{Pa10} gives:
\begin{equation}
  \label{eq:gamma0}
  \gammazerop=-{\gammaeep}-{\gammammp}-
  {\gammaemp}-{\gammamep}=-0.90 \pm 0.08 {\rm (stat)} \pm 0.11 {\rm
    (sys)},
\end{equation}
 in the units, $10^{-4}\;\fm^4$, we use throughout for spin polarisabilities. For further discussion, see Ref.~\cite{Gr12}.


\section{Results}
\label{sec:results}

The  data sets we shall use to fit our free parameters are
discussed extensively in our recent review, Ref.~\cite{Gr12}. For the reasons
explained there, in the low-energy region we include data from
Refs.~\cite{Hy59,Go60,Pu67,Ba74,Ba75,Fe91,Zi92,Ha93,MacG95,OdeL},
 but not from Refs.~\cite{Ox58,Be60}. 
The Baranov 150$^\circ$ data \cite{Ba74,Ba75} were also
excluded.  Above the photoproduction cusp, the data of Hallin~\cite{Ha93}
rises much more strongly than that of \OdL~\cite{OdeL}. Hence the former was excluded from the
analysis we presented in Ref.~\cite{Gr12}. This leaves a gap in the database
between $\wlab=164$ MeV and $\wlab=198$ MeV, so although we quote a cutoff of
170 MeV on the data set for our low-energy fits, in practice we could have
chosen this to be anywhere in the range 164--198 MeV without altering the
results. (For further discussion see Sec.~\ref{sec:data} below.) Below our
cutoff two further points are excluded as clear outliers, namely Federspiel
($135\deg$, 44~MeV) and \OdL\ ($133\deg$, 108~MeV).  For the other \OdL\
points we follow Wissmann \cite{Wissmannmonograph} and add a point-to-point
systematic error of 5\% in quadrature with the statistical error for the \OdL\
data set (not shown in plots). Lastly, although we remove the data of
Hallin~\cite{Ha93} above 150 MeV, the data from this experiment at or below
the photoproduction cusp are completely consistent with the rest of the
database, and so, to maximise our statistical power, we include it in our
fits.

We float the normalisation of each of these data sets within the quoted
normalisation uncertainty, by using the standard augmented $\chi^2$ function as 
given in Eq.~(4.19) of Ref.~\cite{Gr12}; see Ref.~\cite{Baranov:2001} for more details.
 (The MacGibbon data
\cite{MacG95} is counted as two data sets since the higher-energy untagged
data is treated separately). 

For the physical parameters in the $\piN$ sector of the theory we use \cite{PDG}
$\mpi^\pm= 139.6$~MeV, $\fpi= 92.21$~MeV, $\MN=\Mp= 938.3$~MeV, $\ga=
1.27$, $\kappas= -0.22$ and $\kappav= 3.71$.  For the other LECs from $\calL^{(2)}_{\piN}$ we use  $c_1=-0.9$~GeV$^{-1}$, $c_2=0.2$~GeV$^{-1}$ and $c_3=-1.6$~GeV$^{-1}$  
\cite{Be08}. \footnote{Ref.~\cite{Be08} gives
 $ c_1=-0.9^{+0.2}_{-0.5}$, $c_2=3.3 \pm 0.2$, $\ c_3=-4.7^{+1.2}_{-1.0}$, all in GeV$^{-1}$,
from which we subtract the Delta-pole contribution of $\pm4\gpiNDelta^2/9(\MDelta-\MN)$
from $c_2$ and $c_3$ respectively.}
The neutral pion mass
$m_{\pi^0}=134.98$~MeV is used in the $\pi^0$ pole diagram but not elsewhere;
other isospin-breaking effects are neglected. The Delta parameters are fit
to the Breit-Wigner peak and relativistic width as $\MDelta-\MN= 293$~MeV and
$\gpiNDelta =1.425$.

Finally, we remind the reader that we use units of $10^{-4}$~fm$^{3}$ for $\alphae$ and $\betam$, and 
$10^{-4}$~fm$^{4}$ for the spin polarisabilities.

\subsection{Fitting strategy}
\label{sec:fitting}

The aim of this investigation is to extend and merge the approaches of
Hildebrandt et al.\ and Beane et al.\ \cite{Be02, Be04,Hi04}, carrying out a
definitive fit of $\alphae$ and $\betam$ to low-energy data at $\calO(e^2
\delta^4)$ in chiral EFT with a dynamical Delta.  As summarised in
Sec.~\ref{sec:unifiedpc}, the ingredients are as follows.  Born and pion-pole
graphs are included, as are counterterms for $\delta\alphae$ and $\delta\betam$;
see Fig.~\ref{fig:protborn}. In addition we add the $\piN$ and $\pi\Delta$
loops of Figs.~\ref{fig:protloop} and \ref{fig:protdelta}.  The Delta-pole
graphs of Fig.~\ref{fig:protdelta}(i) are treated covariantly, using the
Lagrangian of Eq.~\eqref{eq:PPEMLagrangian} together with the corresponding
expression for the width Eq.~\eqref{eq:deltaself}, and we include
vertex corrections as discussed in Sec.~\ref{sec:gNDvertex},
Fig.~\ref{fig:vertexstructure}. The amplitude is complete to $\calO(e^2
\delta^4)$ in the low-energy region (sometimes referred to below as ``fourth
order'', since it corresponds to $\calO(e^2 P^2=e^2\delta^4)$ for the $\piN$
loops) and $\calO(e^2 \delta^0)$ in the resonance region.

Hildebrandt et al.\ additionally fit the $\gammaN \Delta$ coupling
constant $b_1$ to data from Refs.~\cite{Ha93,OdeL} which satisfied $\w \lsim
240$ MeV. This was done without the inclusion of any width for the
$\Delta(1232)$.  We have examined the advisability of repeating this strategy
(with a Delta width included).  In fits to data below about
$\w=180$~MeV, the 1-$\sigma$-surface in the
$\alphae$-$\betam$-$b_1$ parameter space is large, with a particular degeneracy
between $b_1$ and
$\alphae-\betam$. Extending the fit to higher energies tends to constrain $b_1$, but at the
expense of preferring values of $\alphae$ and $\betam$ which are less compatible
with the low-energy values.  The significance of this finding is hard to judge
in view of the fact that the data set between 170 and 240~MeV is sparse and
inconsistent (as discussed in detail in the review~\cite{Gr12}).

Instead we adopted the following strategy.  It clearly makes sense to use
resonance-region data to determine $b_1$; the cross section there is
essentially proportional to $b_1^4$.  Unlike Hildebrandt et al., our amplitude
is at least NLO in this region, since we use the covariant Delta. This
further has the advantage that the parameters $b_1$ and $b_2$ have the same
meaning as in the photoproduction study of Pascalutsa and
Vanderhaeghen~\cite{PV06}; we can therefore take the ratio $b_2/b_1=-0.34$
from their work. Furthermore, though we choose to treat $b_1$ as a fit
parameter we should get a value which agrees with Ref.~\cite{PV06}
($b_1=3.81$) up to corrections beyond NLO.

However, above 200~MeV the data sets of, on the one hand, Hallin~\cite{Ha93}
and Blanpied~\cite{Blanpied}, and, on the other, Wolf~\cite{Wolf} and
Camen~\cite{Ca02}, together with others working at Mainz~\cite{Mo96,Pe96,Hu97,Wi99}, are
too discrepant for a consistent fit to be obtained. We have chosen to use the
Mainz data for our fits to the higher-energy region, for reasons discussed
further in Ref.~\cite{Gr12}.  Thus we fit $b_1$ to Mainz data between 200 and
325~MeV, then $\alphae$ and $\betam$ to world data below 200~MeV, and iterate
until convergence is reached.  We do not pay too much attention to the quality
of the high-energy fit, since our amplitude is only NLO in this region; in
addition the data are noisy and there are apparent trends in the data that are
not fully reproduced.  We do, however, expect a high quality of fit at low
energies.

In Ref.~\cite{Gr12} we presented an $\calO(e^2 \delta^3)$ fit to the Compton
database using just this strategy (note though that $\gammaN \Delta$
vertex corrections were not included in the amplitude there). The Hallin data
were ultimately excluded because of their unacceptable
$\chi^2$.  
We obtained a good fit with a $\chi^2$ of 106.1 for 124 degrees of freedom
(d.o.f.) (135 data points, 9 floating normalisations and 2 fit parameters).
This calculation with the $\piN$-loops computed to third-order---$\calO(e^2
P)$---in the $\chi$EFT expansion---sets a standard for the calculation we
present here.

However simply repeating such a fit with the $\piN$-loop calculation improved
to fourth order does not yield an acceptable $\chi^2$.  An examination of the
work of Refs.~\cite{Be02, Be04,Hi04} reveals why this is the case: both the
$\calO(e^2 P^2)$ pion loops and the Delta significantly improve agreement
compared to the pure third-order HB$\chi$PT result by raising the cross
section in the cusp region. Their combined effect raises the cross section
above the data. The values of the spin polarisabilities also suggest this
problem. They are significantly closer to the indicative values provided by
dispersion-relation (DR) analyses in the Delta-less theory at fourth order
in HB$\chi$PT than they are at third order. However, adding the Delta-pole
then provides a substantial additional $\gammamm$ contribution \cite{hhkk}.
This suggests a resolution of the problem: just as adding the Delta to the
$\calO(e^2 P)$ Delta-less theory required the promotion of the fourth-order
counterterms for $\alphae$ and $\betam$, so, when fourth-order $\piN$ loops and
an explicit Delta are both considered, one or more of the (nominally
$\calO(e^2 P^3)$) counterterms for the spin polarisabilites needs to be
promoted, and included in our $\calO(e^2 \delta^4)$ calculation.

We have therefore explored the effect of varying each of the spin
polarisabilities in turn, i.e. fitting one spin polarisability, along with
$\alphae$ and $\betam$ to the low-energy data.  (Fits where more than one
spin-polarisability is varied produce soft directions in parameter space which
limit our ability to extract $\alphae$ and $\betam$ with reasonable errors.) We
determine the optimal spin polarisability to include amongst the fit
parameters according to the following selection criteria: the low-energy
$\chi^2$ should be good; the fitted value of the spin polarisability should be
``sensible''---which in practice means bringing $\gammazerop$ closer to $-1$
than with the pure fourth-order fit; the Baldin-constrained fit should be
compatible with that obtained without the constraint; and the reproduction of
the high-energy data should not be badly compromised (for reasons already
explained, we do not rely solely on the $\chi^2$ value to determine this).  We
find that on almost every count varying $\gammamm$ is superior to varying
$\gammaem$, $\gammame$, or $\gammaee$; in particular it gives the lowest
low-energy $\chi^2$.  Furthermore $\gammamm$ is the spin polarisability for
which the pure fourth-order prediction is most discrepant with DR estimates
\cite{Report}, and the only one (with an acceptable $\chi^2$) with a fitted
value closer to the DR estimates than the unfitted one. For all those reasons
we choose to fit $\gammamm$ as an additional parameter when presenting our
final $\calO(e^2 \delta^4)$ result.  We would caution, though, that we do not regard 
our extracted value for this parameter as definitive, since we believe that polarised 
Compton scattering is a much better arena in which to explore the spin polarisabilities.

\subsection{Results with the $\calO(e^2 \delta^4)$ low-energy amplitude}
\label{sec:fits}

We fit both without and with the Baldin-sum-rule constraint
$\alphae+\betam=13.8\pm0.4$ (see Sec.~\ref{sec:constraints}).  The statistical
errors are obtained from the $\chi^2_{\text{min}}+1$ ellipsoid (ellipse).
Since $b_1$ is fit to an independent higher-energy data set, its small
statistical error in principle feeds through to produce an additional small
error on the low-energy parameters, but this is negligible compared to other
uncertainties.

Our result then is, for the fit without the Baldin sum rule constraint:
\begin{equation}
  \alphaep=11.7\pm0.7(\text{stat})\pm0.6(\text{theory}),\qquad
  \betamp =3.8\pm0.55(\text{stat})\pm0.6(\text{theory}) 
  \label{eq:4th-fit-NB}
\end{equation}
with $\gammamm=2.9\pm0.65(\text{stat})$ and $b_1=3.62\pm 0.02(\text{stat})$,
as well as a $\chi^2$ of 110.5 for 134 d.o.f. For the Baldin-constrained fit,
we obtain $\alphae-\betam=7.5\pm0.7(\text{stat})\pm0.6(\text{theory})$
  or
\begin{equation}
  \begin{split}
    \alphaep=&\,10.65\pm0.35(\text{stat})\pm0.2(\text{Baldin})
    \pm0.3(\text{theory}),\\
    \betamp =&\,3.15\mp0.35(\text{stat})\pm0.2(\text{Baldin})
    \mp0.3(\text{theory}),
  \end{split}
  \label{eq:4th-fit-Bal}
\end{equation}
with $\gammamm=2.2\pm0.5(\text{stat})$ and $b_1=3.61\pm 0.02(\text{stat})$,
with a $\chi^2$ of 113.2 for 135 d.o.f.\ If we instead take $b_1=3.8$ from
Ref.~\cite{PV06}, the Baldin-constrained value of $\betam$ drops only slightly
from 3.15 to 3.0, well within the statistical uncertainty.  In both
  parameter sets we have also included estimates of the individual residual
  theoretical uncertainties from higher-order corrections. Before we can
  justify them in Sec.~\ref{sec:orders}, we need to explore  the sensitivity of the fits to choices of the parameters and data
  sets, as well as convergence issues.
  
\begin{figure}[tbp]
  \begin{center}
    \includegraphics*[width=.4\linewidth]{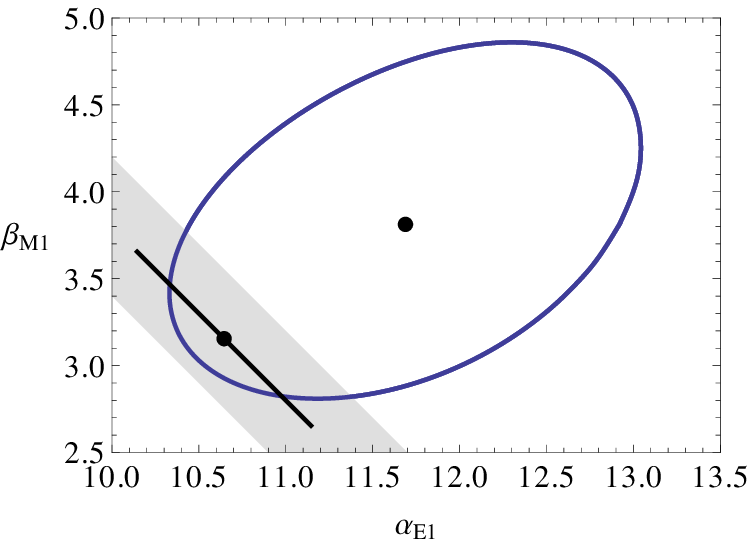}
    \caption {(Colour online) The projections of the $1\,\sigma$ ellipsoid and
      Baldin-constrained ellipse onto the $\alphae$-$\betam$ plane, giving an
      ellipse and a line respectively.  The grey band shows the Baldin
      constraint.}
    \label{fig:chi-4th}
  \end{center}
\end{figure}

The statistical errors are calculated from the boundaries of the $\chi^2_{\text{min}}+1$ ellipsoid or ellipse.
In contrast, our 1\,$\sigma$ regions in 
3d-parameter space are calculated according to $\chi^2=\chi^2_{\text{min}} + 3.5$ (or $\chi^2_{\text{min}} + 2.3$ for 2d), and encompass 68\% of the probability. The projections of such  $1\,\sigma$
regions in  $\alphae$-$\betam$-$\gammamm$ parameter space onto the $\alphae$-$\betam$ plane are shown in
Fig.~\ref{fig:chi-4th}, together with the Baldin-constraint band.  We can see
that the two extractions are fully compatible with one another, as we had
required for an acceptable fit. We have also checked that the $\chi^2$ is not
unacceptably large for any individual data set, and that none of the
normalisations floats beyond the quoted systematic error. In particular, the
65 \OdL\ points are fit with a $\chi^2$ of 69 (unconstrained) and 71
(constrained) with negligible floating of the normalisation.  Our slightly
restricted set of world low-energy data is clearly highly compatible, and
indeed the errors on some sets seem to be overstated.

\begin{figure}[tbp]
  \begin{center}
    \includegraphics*[width=\linewidth]{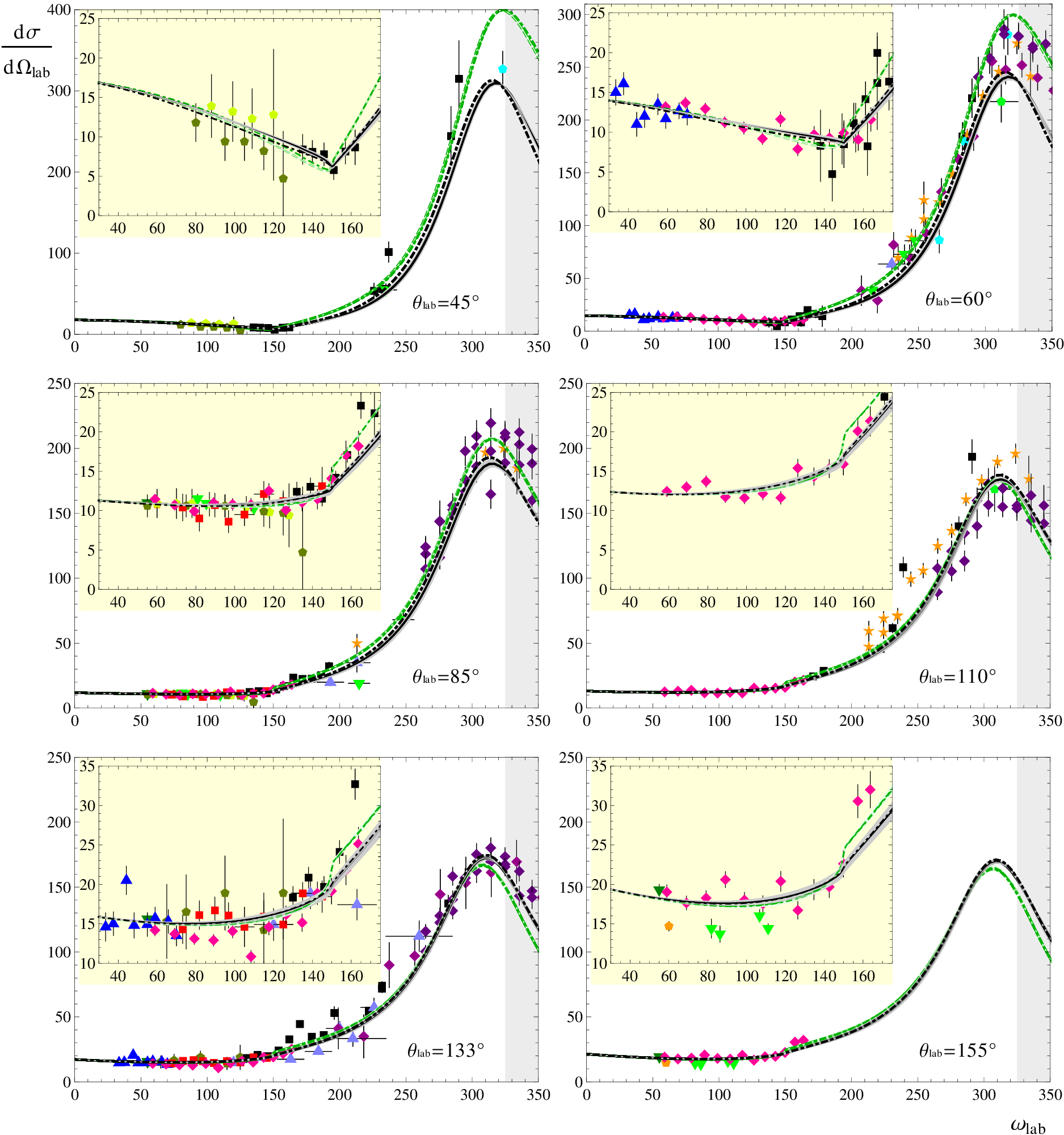}
    \caption {(Colour online) Results of fits to Compton scattering data, as
      described in the text.  The differential cross section in nb/sr is plotted as a
      function of lab energy in MeV, at fixed lab angle. Black and black
      dot-dashed curves are fourth-order Baldin-constrained and unconstrained
      fits respectively (with a band showing the 1 $\sigma$ errors on the
      former); green dashed (with band) and dot-dashed curves are the
      corresponding third-order fits (see Sec.~\ref{sec:orders}).  Parameters
      are given in  Eqs.~\eqref{eq:4th-fit-Bal}, \eqref{eq:4th-fit-NB},
      \eqref{eq:3rd-fit-Bal}, \eqref{eq:3rd-fit-NB} respectively.  The insets
      show the low-energy fit region, and the grey area is beyond the
      high-energy fit region.  The magenta/purple diamonds are MAMI data,
      principally Refs.~\cite{OdeL,Wolf}, while the black squares are the data
      of Hallin~\cite{Ha93}, and the yellow stars are from Blandpied
      Ref.~\cite{Blanpied}. For the definition of other symbols see
      Ref.~\cite{Gr12}. The data shown are within 5$^\circ$ of the nominal
      angle.}
    \label{fig:data-4th}
  \end{center}
\end{figure}

In Fig.~\ref{fig:data-4th} the results are shown (along with third-order fits
close to those of Ref.~\cite{Gr12} which will be discussed later), together
with all the world data, including those data sets that we did not include in
our fits.
As expected, the differences between constrained and unconstrained fits are
small. Essentially no systematic deviation from the trend of the low-energy
data is visible.  The trend of the high-energy Mainz data is well captured,
except for a tendency to fall somewhat low at forward angles.

For reference, with $b_1=3.61$ and $b_2=-0.34b_1$, the other three
spin-polarisabilities take the values $\gammaee=-1.1$, $\gammaem=-0.4$ and
$\gammame=1.9$ (excluding the $\pi^0$ pole contribution which is -45.9 for
$\gammapip$). See table 4.2 of the review \cite{Gr12} for values in other EFTs
and in dispersion-relation extractions.

\subsection{Further details of the fit}
\label{sec:details}
\subsubsection{Sensitivity of the differential cross section to $b_1$,
  $\alphae$, and $\betam$}
\label{sec:sensitivity}

\begin{figure}[tbp]
  \begin{center}
    \includegraphics*[width=\linewidth]{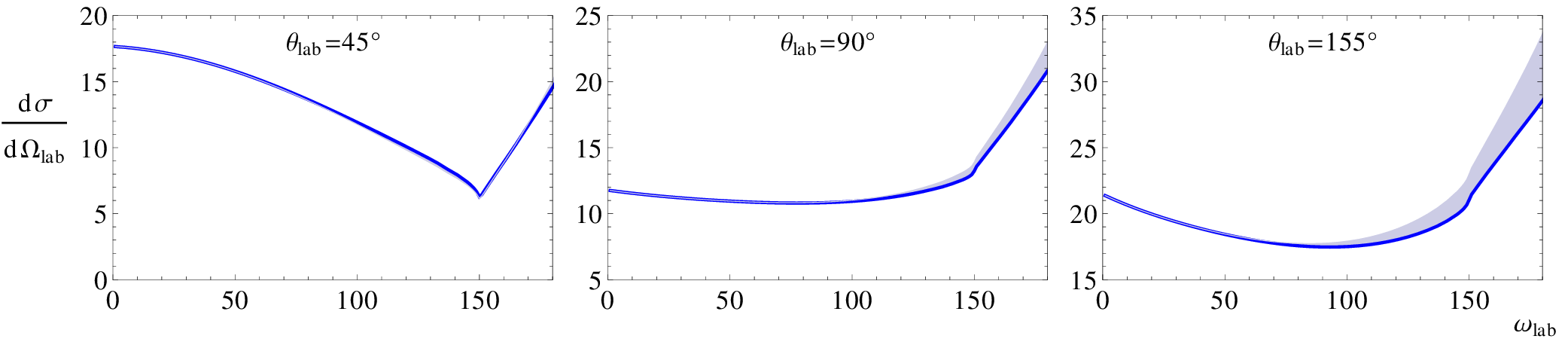}
    \caption {(Colour online) Sensitivity of the fourth-order differential cross section
      (in nb/sr) to $b_1$, plotted as a function of lab energy in MeV at three
      different fixed lab angles, $\thetalab=45^\circ$, $90^\circ$, and
      $155^\circ$. The solid line has $b_1=3.1$, while the lighter band spans
      spans the range to $b_1=4.1$.  Other parameters are those of the
      fourth-order Baldin-constrained fit Eq.~\eqref{eq:4th-fit-Bal}.}
    \label{fig:b1var}
  \end{center}
\end{figure}

In Fig.~\ref{fig:b1var} we show the sensitivity of the low-energy fourth-order
cross section to varying $b_1$ within the rather wide band 3.1-4.1. 
Note that if $\gammamm$ is not fixed, it has a contribution from the Delta
which is proportional to $b_1^2$, and so the variation of the cross section
with $b_1$ is somewhat larger, especially at forward angles. However, the
overall scale of the effect is similar. We observe that there is virtually no
sensitivity to $b_1$ below the cusp at more forward angles, which explains why
Beane et al.\ \cite{Be02, Be04} were able to fit this region so well.

\begin{figure}[tbp]
  \begin{center}
    \includegraphics*[width=\linewidth]{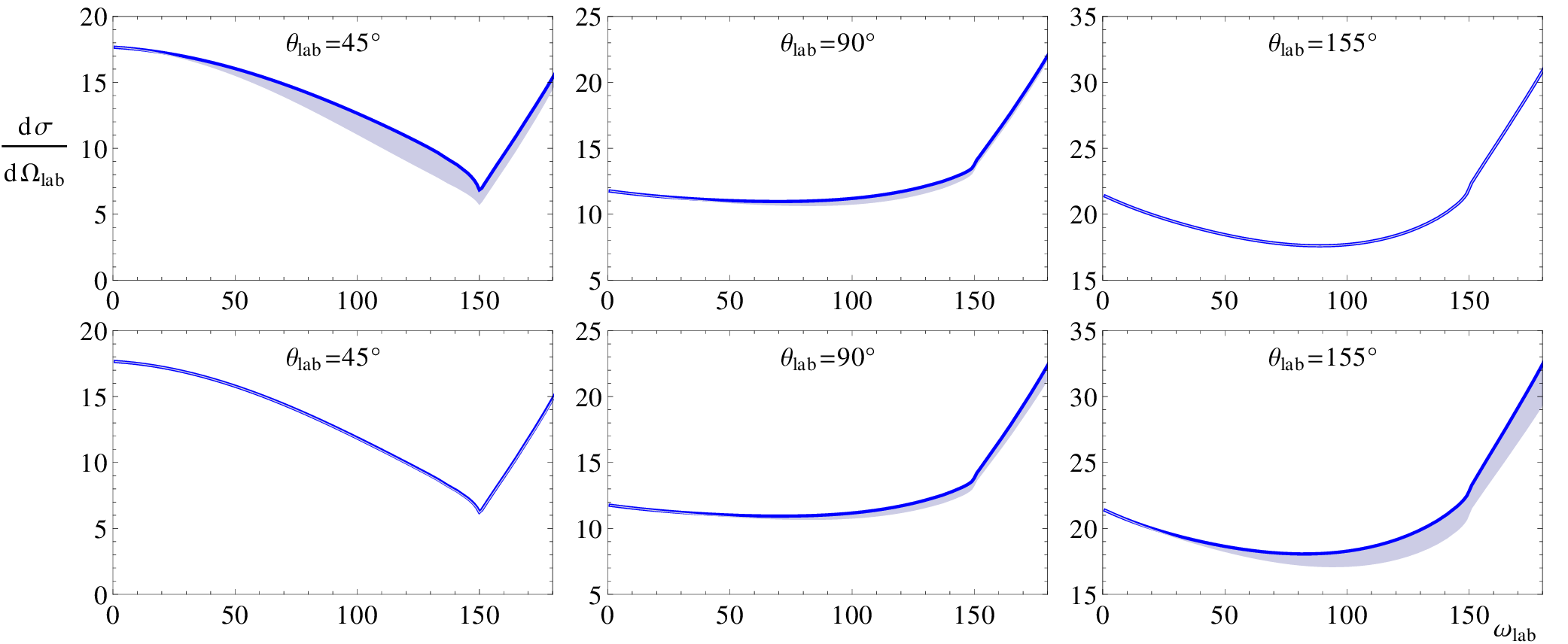}
    \caption {(Colour online) Sensitivity of the fourth-order differential cross section
      (in nb/sr) to $\alphae\pm\betam$, plotted as a function of lab energy in
      MeV at three different fixed lab angles, $\thetalab=45^\circ$,
      $90^\circ$, and $155^\circ$. The upper line shows the sensitivity to
      $\alphae+\betam$; the solid line corresponds to $\alphae+\betam=11.8$, and
      the band spans the range to $15.8$. The lower line shows the sensitivity to $\alphae-\betam$; the solid line
      corresponds to $\alphae-\betam=6$, and the the band spans the range to
      $\alphae-\betam=9$.  Other parameters are those of the
      fourth-order Baldin-constrained fit Eq.~\eqref{eq:4th-fit-Bal}.}
    \label{fig:alphabetavar}
  \end{center}
\end{figure}

In Fig.~\ref{fig:alphabetavar} we show the sensitivity to varying
$\alphae\pm\betam$ within the bands 11.8-15.8 and 6-9 respecttively.  We see, as expected, that most sensitivity to $\alphae+\betam$ ($\alphae-\betam$) occurs at forward (backward) angles. Since
there is no modern data below 60$^\circ$ it is not surprising that---as can be
seen from Fig.~ \ref{fig:chi-4th} and as we will discuss further in
Sec.~\ref{sec:orders}---sensitivity to $\alphae+\betam$ is significantly less
than to $\alphae-\betam$, suggesting that the Baldin-constrained fits are more
reliable.

Further exploration reveals that sensitivity to $\betam$ is almost vanishing at
90$^\circ$ even for energies up to the peak of the resonance.  Of course the
contribution to $A_1$ of Eq.~\eqref{eq:Asinw} vanishes at this angle. However,
there is a piece of $\frac{d \sigma}{d \Omega} \sim |A_2|^2$, and this, in
fact, is a  dominant part at 90$^\circ$ in the resonance region.
Nevertheless, the observation stands.

From Figs.~\ref{fig:b1var} and \ref{fig:alphabetavar} together we can see the
degeneracy between $b_1$ and $\alphae-\betam$ referred to in
Sec.~\ref{sec:fitting}, and hence appreciate the difficulty of fitting both to
the low-energy data alone.  Our strategy of fitting $b_1$ to data above
200~MeV eliminates this problem.

\subsubsection{Choice of data set}
\label{sec:data}
After the elimination of inconsistent data sets, we are left with a choice of
energy to define the top of the ``low-energy'' region.  Accepting that Hallin
data above the cusp cannot be accommodated means that there is no other modern
data between 164 and 198~MeV, and indeed only \OdL\ data between the cusp and
164~MeV. Above 200~MeV not only is the data sparse and inconsistent, but the
prominence of the Delta means that our amplitude, which is only NLO in the
resonance region, is less trustworthy.  So we prefer not to extend our
``low-energy'' fit into this region. On the other hand, since the amplitude
above the cusp is markedly more different between third and fourth order than
that below, a conservative choice of fit region would stop at 150~MeV---in
which region the Hallin data is perfectly consistent. So to test our
sensitivity to the choice of data sets and cut-off, we consider three
possibilities: (I) $\omega_{\text{max}}=150$~MeV, Hallin included (137 points,
10 data sets); (II) $\omega_{\text{max}}=164$~MeV, Hallin excluded (135
points, 9 data sets); and (III) the union of both sets (147 points, 10 data
sets).
\begin{figure}[tbp]
  \begin{center}
    \includegraphics*[width=.4\linewidth]{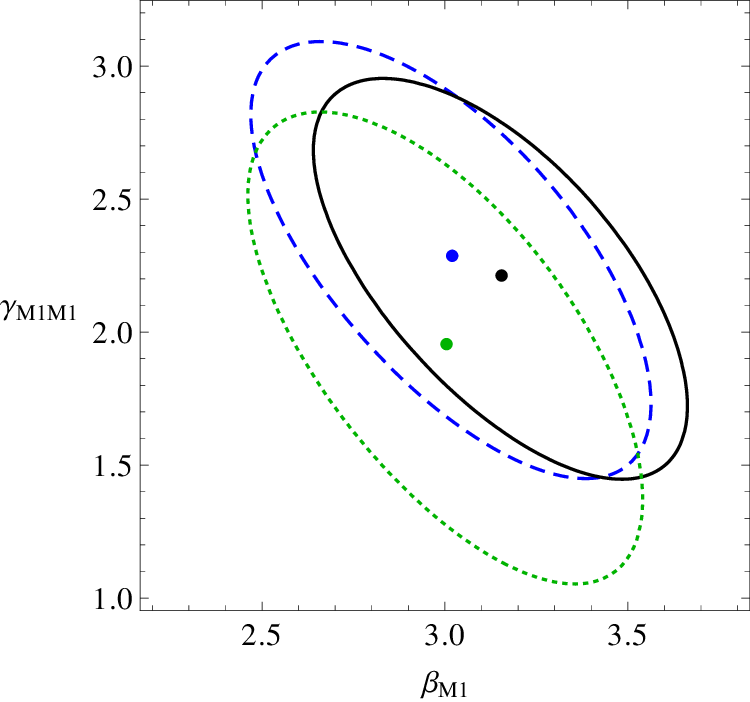}
    \caption {(Colour online) $1\,\sigma$ curves in the $\betam$-$\gammamm$
      plane for Baldin-constrained fits to three choices of data sets
      described in the text: Set (I) green, dotted; Set (II) blue, dashed; and
      Set (III) black, solid.  (III) uses the most data and is our final
      choice.}
    \label{fig:chidatavar}
  \end{center}
\end{figure}

Fig.~\ref{fig:chidatavar} shows that the results from all three strategies are
highly consistent; the values for $\betam$ in the Baldin-constrained fits, with
statistical errors, are (I) $\betam=3.00 \pm 0.36$, (II) $\betam=3.01\pm0.36$,
and (III) $\betam= 3.15 \pm0.34$ where excessive precision has been used simply
to demonstrate the slight differences between the three.  Though it clearly
hardly matters which one we use, we choose as standard the largest data set,
(III).

We also consider the effect of changing the upper cut-off on the higher-energy
data to which we fit $b_1$; a cut-off of 350~MeV favours a somewhat larger
$b_1$ of 3.69, and the Baldin-constrained $\betam$ in the corresponding
low-energy fit drops to 3.08 ($\chi^2=112$), which is clearly well within our
uncertainties.

\subsubsection{The effect of loop corrections to the $\gammaN \Delta$
  vertices}
\label{sec:vertexcompare}
While the the $\piN$ loop vertex corrections of Fig.~\ref{fig:vertexstructure}
have previously been considered in photoproduction \cite{PV06}, this
calculation is the first to include them consistently in a study of Compton
scattering.  Their principal effect is to generate a significant imaginary
part in the vertex above the photoproduction threshold, and hence, since the
Delta-pole contribution to the cross section scales as $|b_1(\omega)|^2$,
to decrease the resonance peak height for a given $b_1$ (recall that
$\mbox{Re}(b_1)$ is fixed to be the real part of the running coupling constant
at resonance).  In practice, this means that when we fit $b_1$, we get a
slighly higher value if the coupling runs than if it does not, somewhat
reducing the slight discrepancy between the values obtained here and in
Ref.~\cite{PV06}. A less significant effect is that the real part also runs,
and we recall that this running is essentially arbitrary since there is a LEC
($D_1$ of Eq.~\eqref{eq:subsubleading}) whose finite part we choose not to include.

\begin{figure}[tbp]
  \begin{center}
    \includegraphics*[width=.7\linewidth]{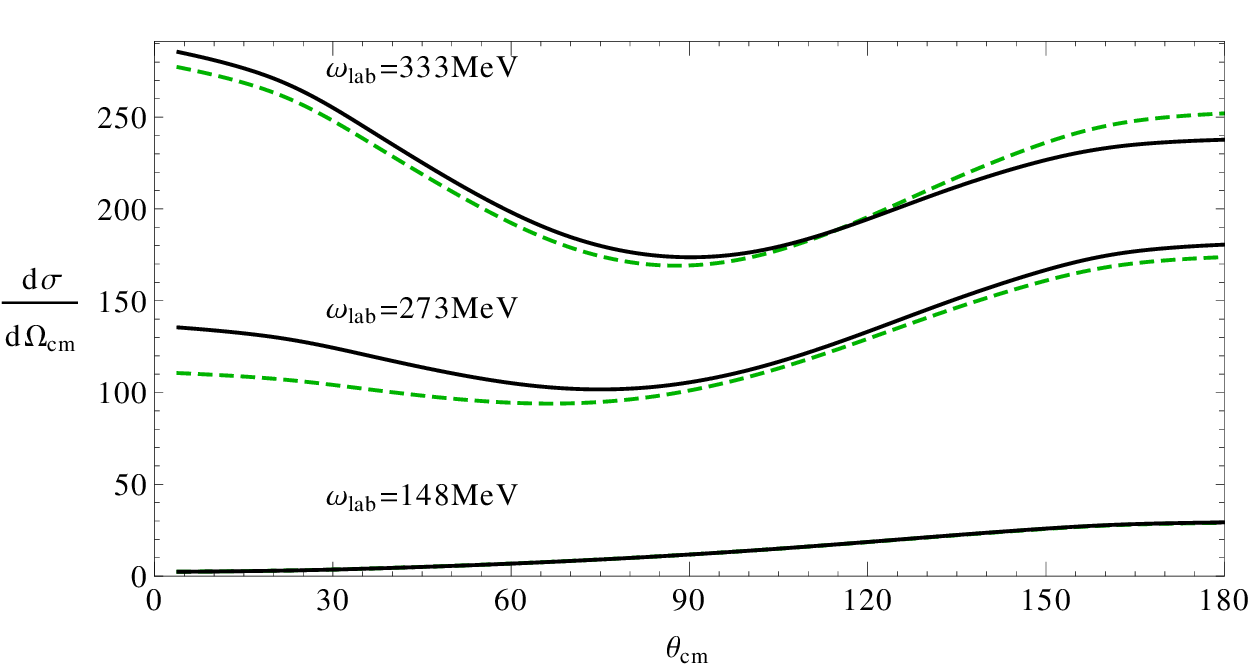}
    \caption{(Colour online) Differential cross sections (nb/sr) as a function
      of c.m. angle with (black, solid) and without (green, dashed) $\piN$
      loop corrections to the $\gammaN \Delta$ vertex at three
      different energies. Best-fit Baldin-constrained parameters are used for
      each (see text).}
    \label{fig:run}
  \end{center}
\end{figure}

If we fit only to data below the cusp (where the imaginary part vanishes), the
$\chi^2$ and $\betam$ in the Baldin-constrained fit differ by less that a
percent with and without running.  Including the rest of the \OdL\ data, the
fit without running has $\chi^2=117$ and $\betam=3.2\pm0.35$ (cf.\ $\chi^2=109$
and $\betam=3.15\pm0.35$ with the running), which is only a very slight change.
As shown in Fig.~\ref{fig:run}, once the low-energy parameters are adjusted,
the curves below the cusp with and without the running are indistinguishable
by eye.  In the resonance region the effects are of course more marked, and
there is a modest improvement in the fit when running is included.

\subsubsection{Convergence and residual theoretical uncertainties}
\label{sec:orders}

\begin{figure}[tbp]
  \begin{center}
    \includegraphics*[width=\linewidth]{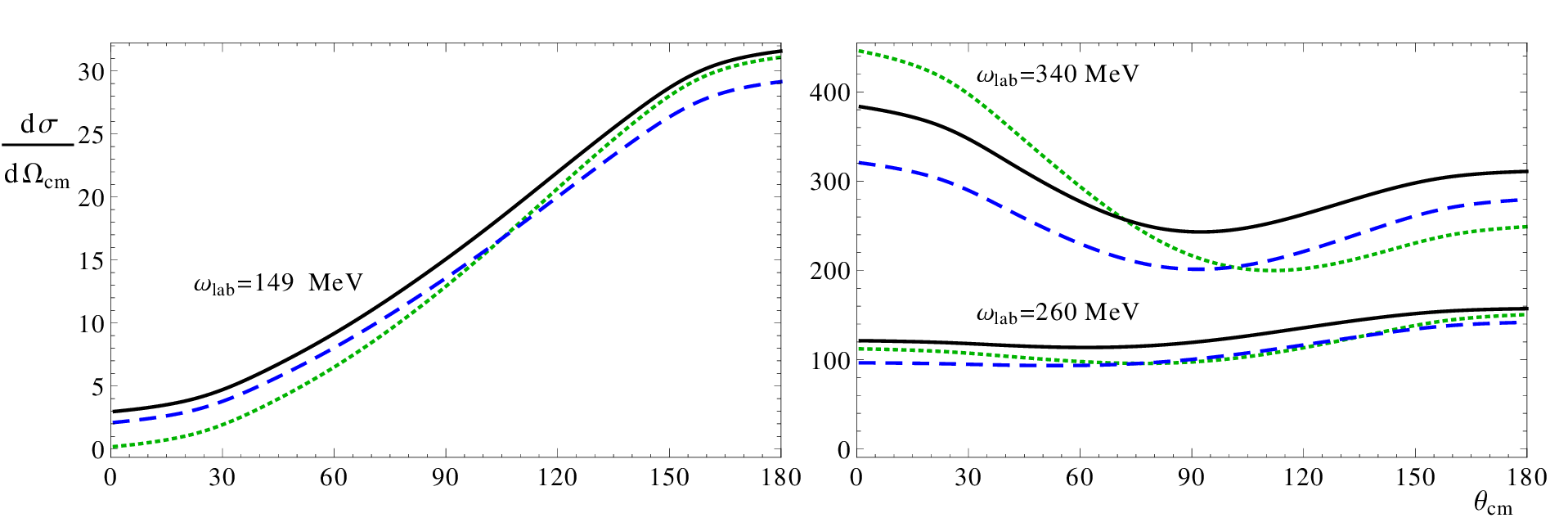}
    \caption {(Colour online) Differential cross sections (in nb/sr) at third order (green,
      dotted), fourth order without 2nd-order LECs (blue, dashed) and full
      fourth order (black, solid), all without $\piN$ loops in
      $\gammaN  \Delta$ vertices and with $\alphae=10.8$, $\betam=3$,
      $b_1=3.8$ as well as unadjusted $\gammamm$. Curves at $\wlab=149$ MeV
      (left panel) and $\wlab=260$ and $340$ MeV (right panel) are shown as a
      function of c.m.~angle.}
    \label{fig:magmom}
  \end{center}
\end{figure}

We are now in a position to examine the convergence of our results as we go
from $\calO(e^2\delta^3)$ (``third order'') to $\calO(e^2\delta^4)$ (``fourth
order'') in the low-energy amplitude.  The nucleon Born contribution changes
appreciably between these two, whereas the difference between the fourth-order
result and the fully covariant amplitude is negligible, so we choose to use
the fourth-order Born for both amplitudes, just as we use the covariant
Delta-pole graphs. (Since the fourth-order Born amplitude has
polarisability-like terms, the main effect of this is to avoid a spurious
shift in polarisabilities between orders.) Hence the only difference between
the two orders (apart from the vertex corrections) is the graphs of 
Fig.~\ref{fig:protloop}(ii).  These are of two
kinds: $1/\MN$ corrections to the graphs of Fig.~\ref{fig:protloop}(i), which
would be considered part of the third-order calculation in a covariant
formulation, and contributions from the second-order LECs $c_i$ and $\kappas$,
$\kappav$ which are fourth-order in both heavy-baryon and covariant
formulations.  Fig.~\ref{fig:magmom} shows the comparison between third order,
fourth order with the LECs set to zero, and full fourth order, all with the
same values of the polarisabilities $\alphae=10.8$, $\betam=3$. (Of course we do
not switch off $\kappa$ in the Born terms in the first two calculations, only
in the loops.)
We do not show the effects of the $c_i$'s and $\kappa$ separately, as the
former is almost completely negligible. (Recall that in the theory with an
explicit Delta the $c_i$'s have to be readjusted, and are all rather small.)
On the other hand $\kappa$ has an appreciable effect (largely due to
$\kappav$; $\kappas$ is small). The $1/\MN$ convergence is very satisfactory
up to about 300~MeV, with the difference between the third and fourth orders
only exceeding 10\% where the third order is anomalously low, i.e. for forward
angles at the cusp.  We also see that the effect of $\kappa$ is, in the same
range, of the same size as that of $1/\MN$ corrections. Hence there seems
little to be gained in using a covariant formulation for the loops without
including anomalous-magnetic-moment effects in the loop calculation too.
Unsurprisingly, the situation is different in the resonance region, where the
$1/\MN$ corrections are becoming large.

This convergence is very satisfactory from a formal perspective. However,
quite a lot of the low-energy data lie in the region where the difference
between the third- and fourth-order  amplitudes is
accidentally enhanced, and so identical results when one fits the
polarisabilities to the data are not guaranteed.  To test the convergence of
the fourth-order results of section \ref{sec:fits} we perform a fit using the
third-order amplitude, obtaining the following results:
\begin{equation}
  \alphaep=10.2\pm0.5(\text{stat})\pm0.8(\text{theory}),\qquad
  \betamp =2.5\pm0.55(\text{stat})\pm0.8(\text{theory}),
  \label{eq:3rd-fit-NB}
\end{equation}
with $b_1=3.69\pm0.02(\text{stat})$ and $\chi^2$ of 118.6 for 134 d.o.f., and for
the Baldin-constrained fit
\begin{equation}
  \begin{split}
    \alphaep=&\,10.7\pm0.3(\text{stat})\pm0.2(\text{Baldin})\pm0.8(\text{theory}),\\
    \betamp =&\,3.1\mp0.3(\text{stat})\pm0.2(\text{Baldin}),
    \mp0.8(\text{theory}),
  \end{split}
  \label{eq:3rd-fit-Bal}
\end{equation}
with $b_1=3.69\pm0.02(\text{stat})$ and $\chi^2$ of 120.3 for 135 d.o.f.

The total $\chi^2$ is distinctly higher than for the fourth-order fit of
Eqs.~\eqref{eq:4th-fit-Bal}, \eqref{eq:4th-fit-NB} but entirely acceptable; the
preferred normalisation shift for the Hallin data is slightly greater than the
systematic error of 4\%, but omitting the data set has little effect on the
fit parameters.  This fit is shown together with the fourth-order one in
Fig.~\ref{fig:data-4th}.  In the review \cite{Gr12} we performed third-order
fits without running couplings and excluding all Hallin data; for the
Baldin-constrained fit identical values were obtained, while for the
unconstrained fit the results were $\alpha=10.5\pm0.5(\text{stat})$ and
$\betam=2.7\pm0.5(\text{stat})$.
\begin{figure}[tbp]
  \begin{center}
    \includegraphics*[width=.6\linewidth]{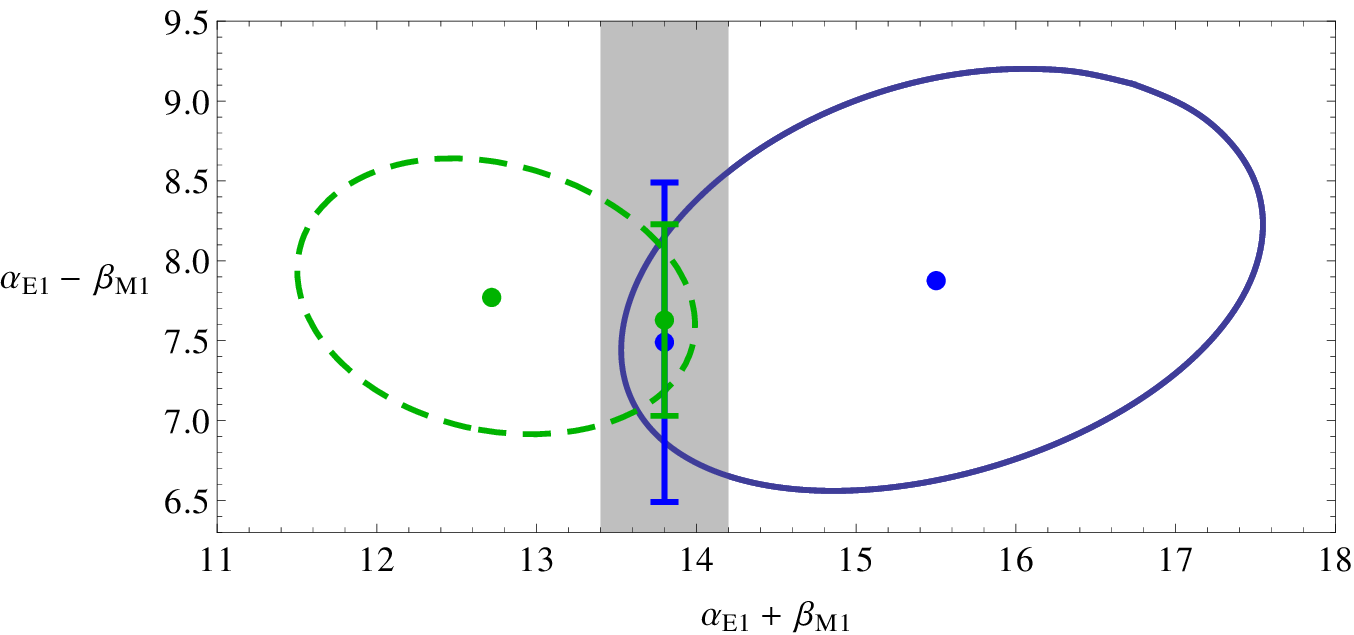}
    \caption {(Colour online) $1\,\sigma$ regions of parameter space for the
      third- and fourth-order, Baldin constrained and unconstrained fits.  The
      green, dashed ellipse and the shorter error bar correspond to third
      order; the blue, solid ellipse and larger error bar are projections of
      fourth-order fits in which $\gammamm$ is also varied, which is why the
      regions are larger. The fourth-order regions are the same as in
      Fig.~\ref{fig:chi-4th}.}
    \label{fig:chi-square3-4}
  \end{center}
\end{figure}

The $1\,\sigma$ curves for the third- and fourth-order fits together are shown
in Fig.~\ref{fig:chi-square3-4}.  We see that the results are fully in
agreement with one another. The range of $\alphae + \betam$ accommodated in
each fit is quite large, and they only just overlap at the 1-$\sigma$ level.
This can be explained by the fact, noted earlier, that the two cross sections
differ most at forward angles near the cusp, which is also where the
sensitivity to $\alphae+\betam$ is greatest, but there is little data at angles
$\theta_{\mathrm lab} < 60^\circ$.

In contrast, the central values of $\alphae-\betam$ in both fits are extremely
close, and neither changes appreciably when the Baldin constraint is imposed.
In view of this, we prefer to quote the fourth-order Baldin-constrained values
of Eq.~\eqref{eq:4th-fit-Bal} as our final result.

Finally, we can now return to the issue of the theoretical accuracy of
  the extractions at $\calO(e^2\delta^4)$ in Eqs.~\eqref{eq:4th-fit-NB}
  and~\eqref{eq:4th-fit-Bal} in Sec.~\ref{sec:fits}.
  Since the polarisabilities first enter at
  $\calO(e^2\delta^2)$ with the parameter-free prediction of Eq.~\eqref{eq:BKM}, 
  we expect corrections to be \emph{a priori} of order
  $\delta^3\sim6\%$ relative to the LO result.  With an average value
  $(\alphaep^\mathrm{LO}+\betamp^\mathrm{LO})/2\approx 7$ to set the scale and $\delta\sim0.4$,
  this leads to an error of $\pm0.5$ (rounded up) for the individual polarisabilities of the free
  fit, or for the combination $\alphae-\betam$ of the Baldin-constrained fit.

To validate this naive dimensional estimate, we can check the
  convergence pattern from order to order, which is predicted to be $\delta\times 7\sim3$ 
 from  LO to NLO and $\delta^2\times 7\sim 1.1$ from  LO to NLO.
 For the free fit, $\alphae$ changes from LO to NLO
  by $2.3$ units, and from NLO to \NXLO{2} by $1.5$ units, while $\betam$
  changes in each case by $1.3$ units, respectively. These are quite compatible
  with the power counting, and suggest an error on the fourth-order fit values of 
 $\pm\delta\times 1.4\sim\pm0.6$.  To be conservative, this is what we quote on our non-Baldin-constrained fit~\eqref{eq:4th-fit-NB}.

However, Fig.~\ref{fig:chi-square3-4} shows that the convergence for 
$\alphae-\betam$ seems to be much better than for  $\alphae+\betam$, the latter
being poorly constrained by the data. Judging by the shift from third to fourth order, 
the error on $\alphae+\betam$ could be taken to be as large as 
$\pm\delta\times 3\sim\pm1.2$, while that on  $\alphae-\betam$ would be neglible. 
To estimate the error on the latter, therefore, we prefer to take $\delta^2$ times
  the LO-to-NLO change of $3.5$, giving $0.6$, and hence an error of $\pm0.3$ on 
the Baldin-constrained values of $\alphae$ and $\betam$ separately.  This also implies that the theory uncertainties for the free fit of  $\alphae$ and $\betam$  are highly correlated. 

A further way of estimating errors in an EFT is to look at the effect of the inclusion
of partial higher-order terms.  We have done this in the low energy region when we include the 
$\gammaN\Delta$ vertex corrections of Section \ref{sec:gNDvertex}. The results for $\alphae$ 
and $\betam$ with and without these corrections differ typically by about $\pm0.1$.  From 
Fig.~\ref{fig:magmom} we might speculate that $\piN$ loops with two insertions of $\kappav$ may in
fact be the dominant correction at \NXLO{3}, but still this is a reassuring result which 
suggests that our theory errors are not underestimated.

For the errors on the third-order results of Eqs.~\eqref{eq:3rd-fit-NB} and 
\eqref{eq:3rd-fit-Bal}, we take the $\pm0.8$ as estimated in the review \cite{Gr12}, values 
which are entirely compatible with the current analysis though almost certainly too large for the Baldin-constrained results.
 
We close by pointing out that the statistical and residual theoretical
  uncertainties in extracting proton polarisabilities are now compatible.

\section{Summary and Outlook}
\label{sec:conclusion}

In this work we have presented a chiral effective theory analysis of $\gammaonp$ scattering data. We employed an EFT with explicit nucleon, pion, and Delta 
degrees of freedom, arranged according to an expansion in the small parameter $\delta \equiv \mpi/(\MDelta-M_N)
\equiv (\MDelta - M_N)/\Lambda$~\cite{PP03}. 

The $\chi$EFT amplitude accurate up to $\mathcal{O}(e^2 \delta^4)$ in the low-energy ($\omega \sim m_\pi$) region was used (where the leading, Thomson, amplitude is $\calO(e^2\delta^0)$). This includes the fourth-order $\piN$ loops of Ref.~\cite{McG01}, 
the $\pi \Delta$ loops first computed 
in  Ref.~\cite{He97}, as well as Delta-pole graphs. At this order short-distance pieces of the proton scalar polarisabilities,
$\alphaep$ and $\betamp$, appear in the Compton amplitude and these, together with the $\gammaN\Delta$ M1 transition strength, $b_1$, are the
 free parameters in our $\chi$EFT description of $\gammaonp$ scattering for $\omega \sim \mpi$. However, instead of fitting $b_1$ in this low-energy region,
 we determine it using data up to $\wlab=325$ MeV and a $\chi$EFT amplitude which includes all $\mathcal{O}(e^2 \delta^0)$ (next-to-leading order) effects
 in the resonance region. Meanwhile, $\alphaep$ and $\betamp$ were fit to a statistically consistent Compton database that covers the domain $0 \leq \wlab \leq 170$ MeV. Fits to this database and to resonance-region data 
were iterated until convergence was achieved. This does not, however, result in an accurate description of data in the low-energy region. 
 By including the short-distance piece of the proton spin polarisability, $\gammamm$---nominally a higher-order effect in $\chi$EFT---in our $\mathcal{O}(e^2 \delta^4)$ amplitude we 
 obtain an excellent fit to the low-energy data. $\gammamm$ is the only spin polarisability where such promotion yields a fit that is consistent with other constraints,
 and with the $\mathcal{O}(e^2 \delta^3)$ fit presented in Ref.~\cite{Gr12}. Since there is little $\gammaonp$ data at forward angles, and it is there that the differential cross section is most sensitive to $\alphaep + \betamp$, we use the Baldin-sum-rule evaluation $\alphaep + \betamp=(13.8 \pm 0.4) \times 10^{-4}~{\rm fm}^3$ from Ref.~\cite{OdeL} to constrain this combination of polarisabilities.
 
 The resulting fit has a $\chi^2$ of 113.2 for 135 d.o.f. We find, in the canonical units of $10^{-4}~{\rm fm}^3$,
  \begin{align}
    \alphaep   = &\,10.65\pm0.35(\text{stat})\pm0.2(\text{Baldin})
        \pm0.3(\text{theory}),\nonumber \\
    \betamp = &\,3.15\mp0.35(\text{stat})\pm0.2(\text{Baldin})
    \mp0.3(\text{theory}),
  \label{eq:4th-fit-Bal-2}
\end{align}
with $\gammamm=[2.2\pm0.5(\text{stat})] \times 10^{-4}~{\rm fm}^4$ and $b_1=3.61\pm 0.02(\text{stat})$. The result for $\alphaep-\betamp$ is very close to that extracted using the $\mathcal{O}(e^2 \delta^3)$ amplitude, and is stable under modifications of the $\gammaonp$ database and the inclusion of certain higher-order corrections to 
the $\gammaN\Delta$ vertex. Importantly, the value of $b_1$ obtained in this fit is consistent with that found in the pion electroproduction study of Ref.~\cite{PV06} at the expected level of
accuracy of both calculations (NLO). 

Eq.~(\ref{eq:4th-fit-Bal-2}) presents a considerably larger value of $\betamp$ than has been found in dispersion-relation extractions of proton polarisabilites from a similar database (see, e.g., Ref.~\cite{Report}). Since $\alphaep - \betamp$ is a free parameter in dispersion-relation fits, it would be interesting to know if such fits can tolerate the values of $\betamp$ found here---and, if they cannot, what additional physics in the dispersion-relation calculation forces $\betamp$ to smaller values. We observe that a $\mathcal{O}(e^2 \delta^3)$ calculation in a variant of EFT that does not use the heavy-baryon expansion predicts $\betamp=(4.0 \pm 0.7) \times 10^{-4}~{\rm fm}^3$~\cite{LP09,LP10}. This is consistent with our extraction (\ref{eq:4th-fit-Bal-2}), but disagrees with the dispersion-relation results in Refs.~\cite{Report,OdeL}.

The $\chi$EFT calculation in the resonance region could be improved to NNLO by including further corrections to the $\gammaN\Delta$ vertex and computing the pertinent pieces of the two-loop self-energy of the Delta in $\chi$EFT~\cite{LvK09}. This would be particularly interesting for the development of a consistent $\chi$EFT description of pion-nucleon scattering, pion photoproduction, and Compton scattering in the resonance region.  The Baldin sum-rule value obtained in Ref.~\cite{OdeL} should perhaps also be re-assessed, since slightly different values have been obtained by other groups~\cite{Ba97,LL00}. The uncertainty on the value of $\alphaep + \betamp$ obtained in this way is on the verge of becoming a significant component in the overall uncertainty in $\alphaep$ and $\betamp$. Meanwhile, 
if the amplitude in the low-energy region were to be improved, a calculation beyond $\mathcal{O}(e^2 \delta^4)$ in the $\delta$ expansion would have to be carried out, and we note that graphs with two $\pi$N loops enter at $\mathcal{O}(e^2 \delta^6)$. As the result (\ref{eq:4th-fit-Bal-2}) makes clear, at this stage the theory error is slightly smaller than the statistical error, and so it seems that the accuracy with which $\alphaep$ and $\betamp$ are known can be more profitably improved by focusing effort in other areas.

In particular, if more data were available between $\pi$N threshold and the $\Delta(1232)$ resonance then we could extend the low-energy fit region and obtain tighter constraints on $\alphaep$ and $\betamp$. The data that exist here are sparse, and different experiments are not consistent with one another. In consequence our database has a sizable gap between $\wlab=164$ MeV and $\wlab=198$ MeV. Data of high precision in this kinematic domain, with well-documented systematic uncertainties, could help reduce the statistical errors on $\alphaep-\betamp$. Indeed, if such experiments extended into the resonance region they might help to discriminate between the conflicting results of, on the one hand, Hallin~\cite{Ha93}
and Blanpied~\cite{Blanpied}, and, on the other, Wolf~\cite{Wolf},
Camen~\cite{Ca02}, and other groups at Mainz~\cite{Mo96,Pe96,Hu97,Wi99}.

Ongoing experiments  to measure $\gammaonp$ asymmetries at $\omega=250$--$350$ MeV (MAMI, Mainz) and $\omega \approx 100$ MeV (HI$\gamma$S, TUNL),  and thereby extract spin polarisabilities, will provide additional testing grounds for our approach~\cite{Mi12,Mi09}. Meanwhile, new data on absolute cross sections from 120--180 MeV will supplement the current database, and, if sufficiently precise, aid the extraction of $\alphaep - \betamp$~\cite{Ho12}.

In fact, the $\chi$EFT amplitude developed here is highly successful in reproducing extant Compton differential-cross-section data. Perusal of Fig.~\ref{fig:data-4th} shows that it describes experiment over a wide range of energies and angles. This is possible because the amplitude contains the key non-analyticities in $\gammaonp$ scattering below $\wlab=350$ MeV: the pion photoproduction cusp  and the $\Delta(1232)$ resonance pole. This benefit comes at a price, though: the chiral dynamics encoded in these non-analyticities describes data so well that effects due to short-distance physics play a relatively small role. Consequently, we had to exercise significant care in the treatment of (sometimes discrepant) data, and also consistently incorporate a number of subtle effects in the $\gamma$p amplitude to obtain accurate values for $\alphaep$ and $\betamp$.

\section*{Acknowledgments}
DRP thanks the Theoretical Physics Group at the University of Manchester for
their hospitality during his time that much of this work was done.
We also acknowledge the organisers and participants of the INT workshop
08-39W: ``Soft Photons and Light Nuclei'' and of the INT programme 10-01:
``Simulations and Symmetries'', both of which also provided financial support.
We are grateful to Jerry Feldman, Vladimir Pascalutsa and Mike Birse for useful discussions.
This work has been supported in part by UK Science and Technology Facilities
Council grants ST/F012047/1, ST/J000159/1 (JMcG) and ST/F006861/1
(DRP), by the US Department of Energy under
grants DE-FG02-93ER-40756 (DRP) and DE-FG02-95ER-40907 (HWG), by US National
Science Foundation \textsc{Career} award PHY-0645498 (HWG), by the Deutsche
Forschungsgemeinschaft and the National Natural Science Foundation of China
through funds provided to the Sino-German CRC 110 ``Symmetries and the
Emergence of Structure in QCD'' (HWG), and by University Facilitating Funds of the
George Washington University (HWG).

\appendix

\section{Details of the amplitudes}
\label{sec:appendix}

The nucleon Born diagrams
calculated to fourth order are as given by McGovern~\cite{McG01}; they are
indistinguishable from the full relativistic form given by Babusci et~al.\
\cite{Ba98} for the energies of interest.  They are obtained from the diagrams of Fig.~\ref{fig:protborn}; the third-order vertex represented by a triangle employs terms given in  Eqs.~(3.8) and (3.9) of Ref.~\cite{FMMS}; the divergences represented by $d_{24}$ and $d_{31}$ in the latter renormalise $Z_{\mathrm N}$  in Fig.~\ref{fig:protloop}(ii)(k) and  the loop in (m).

The fourth-order vertex represented by a diamond in Fig.~\ref{fig:protborn}(iii)(b) contributes to both the $\calO(e^2 P^2)$ Born amplitude and to the polarisabilities.
The LECs $\delta\alphae$ and $\delta\betam$ of Eq.~\eqref{eq:LpiN4} result from
linear combinations of the fourth-order operators $O_{89}$ to $O_{94}$,
$O_{117}$ and $O_{118}$ of Ref.~\cite{FMMS}.
The Born amplitudes also involve fixed-coefficient parts of these
(table 6 of~\cite{FMMS}) together with those of the operators $X_{41}$, $X_{53}$ and
$Y_{11}$.  The relevant $\beta$-functions are those of $O_{16}$ to $O_{20}$,
$O_{152}$, $O_{153}$, $O_{160}$ and $O_{198}$ of Ref.~\cite{MMS} (though there would seem to be a misprint in one or more of the ``eye-graph" contributions as the divergence cancellation for $\delta\alphae$ does not quite work.)

The pion-pole
contributions are given by Bernard et~al.\ \cite{BKM}; this form
does not change at fourth order and is indeed relativistically invariant,
although at fifth order form factors will enter at the vertices, and the LEC $d_{18}$ shifts $\ga/\fpi$ to the numerically almost identical $\gpiNN/\MN$.

The $\piN$ loops of Fig.~\ref{fig:protloop} are given at third order by Bernard
et~al.~\cite{BKM} and at fourth order by McGovern~\cite{McG01}.  At
both third and fourth order, we use $\omega_s=\sqrt{s}-\MN$ to shift the
threshold to the correct energy, as detailed in Sec.~\ref{sec:thresholds}.  The $\pi\Delta$ loops are given by
Hildebrandt et~al.~\cite{Hi04} in their Appendix B, except that we
use the Breit-frame photon energy $\w$ throughout in place of $\omega_s$ and $\omega_u$ to preserve crossing symmetry.  

The $s$- and $u$-channel Delta-pole diagrams are
calculated using the Lagrangian of Pascalutsa and Phillips
\cite{PP03}.
The expressions for the Delta-pole amplitudes given in  Eqs~(A3) and (A4)  of that paper
refer to a redundant, covariant set of eight operators. We note that the correct forms of three of these are
\begin{equation}
\calO_5^{\mu\nu}={q}^\mu{q'}_\alpha\gamma^{\alpha\nu}+\gamma^{\mu\alpha}{q}_\alpha{q'}^\nu,\quad\calO_6^{\mu\nu}={q}^\mu{q}_\alpha\gamma^{\alpha\nu}+\gamma^{\mu\alpha}{q'}_\alpha{q'}^\nu,\qquad
\calO_8^{\mu\nu}=\ii\epsilon^{\mu\nu\alpha\beta} {q}_\alpha{q'}_\beta\,;
\end{equation}
 the others are given in their Eq.~(51).
Other misprints in Ref.~\cite{PP03} are noted in Ref.~\cite{Gr12}; in particular 
the expression for the width given in their Eq.~(42) is a misprint for our Eq.~\eqref{eq:deltaself}.
These eight operators may be written in terms of the six Breit frame operators of Eq.~\eqref{eq:Asinw}, which we will denote ${\bf t}_i$ (with e.g.\ ${\bf t}_1=\vec{\epsilon}\,'^*\cdot \vec{\epsilon}$) using
\begin{equation}
\epsilon'^*_\mu\calO_i^{\mu\nu}\epsilon_\nu=2\MN \sum_{j=1}^6 C_{ij} {\mathbf t}_j,
\end{equation}
where
\begin{equation}\small
\renewcommand*{\arraystretch}{1.5}
{\bf C}=\left(
\begin{array}{cccccc}
 \frac{E}{\MN } & 0 & 0 & 0 & 0 & 0 \\
 0 & \frac{\w^2 E}{\MN } & 0 & 0 & 0 & 0 \\
 0 & 0 & 1-\frac{\w^2 (z-1)}{2 \MN  (\MN +E)} & 0 & \frac{\w^2}{4 \MN  (\MN +E)} &
   -\frac{\w^2}{4 \MN  (\MN +E)} \\
 \frac{\w^3 (z-1)}{\MN } & 0 & 0 & \w^2 & 0 & 0 \\
 0 & \frac{\w^3}{\MN } & -\frac{\w^4 \left(z^2-1\right)}{2 \MN  (\MN +E)} & 0 &
   \w^2+\frac{(z+1) \w^4}{4 \MN  (\MN +E)} & -\frac{\w^4 (z+1)}{4 \MN  (\MN +E)}
   \\
 0 & \frac{\w^3}{\MN } & -\frac{\w^4 \left(z^2-1\right)}{2 \MN  (\MN +E)} & 0 &
   \frac{\w^4 (z+1)}{4 \MN  (\MN +E)} & \w^2-\frac{\w^4 (z+1)}{4 \MN 
   (\MN +E)} \\
 0 & -\frac{\w^5 (z-1)}{\MN } & \w^4 \left(z^2-1\right) & 0 & -\w^4 z & \w^4 \\
 0 & 0 & -\frac{\w^3 (z-1)}{\MN } & 0 & \frac{\w^3}{2 \MN } & -\frac{\w^3}{2 \MN }
\end{array}
\right),
\end{equation}
 $E$ is the Breit-frame nucleon energy, $E=\sqrt{\MN^2-t/4}$, and 
$z=\cos\theta=t/(2\w^2)+1$.



\begin{thebibliography}{99}


\bibitem{Gr12} H.~W.~Grie\ss hammer, J.~A.~McGovern, D.~R.~Phillips and
  G.~Feldman, Prog.\ Part.\ Nucl.\ Phys.\  {\bf 67}  (2012) 841. 

\bibitem{Report} D.~Drechsel, B.~Pasquini and M.~Vanderhaeghen,
  Phys.\ Rept.\ {\bf 378}  (2003) 99.

\bibitem{Sc05} M.~Schumacher,
  Prog.\ Part.\ Nucl.\ Phys.\ {\bf 55} (2005) 567.
 
 \bibitem{Ox58} C.~L.~Oxley,
  Phys.\ Rev.\ {\bf 110}  (1958) 733.
  
\bibitem{Hy59} L.~G.~Hyman, R.~Ely, D.~H.~Frisch and M.~A.~Wahlig,
  Phys.\ Rev.\ Lett.\ {\bf 3} (1959) 93.

\bibitem{Go60} V.~I.~Gol'danskii {\it et al.}, Sov. Phys. JETP  {\bf 11} (1960) 1223.

  \bibitem{Be60} G.~Bernardini {\it et al.} Nuovo Cimento {\bf 18} (1960) 1203.
  
\bibitem{Pu67} G.~Pugh {\it et al.}, Phys. Rev. {\bf 105} (1957) 982 and
  references therein; MIT Summer study 1967 p.\ 555.

\bibitem{Ba74} P.~S.~Baranov {\it et al.}, Phys. \ Lett. B {\bf 52} (1974) 122.

\bibitem{Ba75} P.~S.~Baranov {\it et al.}, Sov. J. Nucl. Phys. {\bf 21}  (1975) 355.

\bibitem{De61} J. W. DeWire, M. Feldmann, V. L. Highland, R. Littauer
Phys. Rev. {\bf 124} (1961) 909.  

\bibitem{Ba66a} P. S. Baranov, L. I. Slovokhotov, G. A. Sokol and L. N. Shtarkov,
Sov.\ Phys.\ JETP {\bf 23} (1966) 242.

\bibitem{Ba66b} P. S. Baranov et al,
Sov.\ J. Nucl.\ Phys.\ {\bf 3} (1966) 791.

\bibitem{Gr67} 
  E.~R.~Gray and A.~O.~Hanson,
  Phys.\ Rev.\  {\bf 160} (1967) 1212.

\bibitem{Ge76} H.~Genzel, M.~Jung, R.~Wedemeyer {\it et al.},
    Z.\ Phys.\ {\bf A279} (1976) 399.

 \bibitem{Weinberg79} S.~Weinberg,
  Physica A {\bf 96} (1979) 327.


\bibitem{GL82} J.~Gasser and H.~Leutwyler,
  Phys.\ Rept.\ {\bf 87}  (1982) 77.


\bibitem{Gasser:1983yg} J.~Gasser and H.~Leutwyler,
  Annals Phys.\ {\bf 158} (1984) 142.

\bibitem{JM91a} E.~E.~Jenkins and A.~V.~Manohar, lectures at ``Workshop on
  Effective Field Theories of the Standard Model", Dobogoko, Hungary, Aug
  1991.  In ``Dobogokoe 1991, Proceedings, Effective field theories of the
  standard model", pp. 113-137.
 \bibitem{JM91b}  E.~E.~Jenkins and A.~V.~Manohar,
  Phys.\ Lett.\ B {\bf 259} (1991) 353.
  
    \bibitem{BKM} V.~Bernard, N.~Kaiser and U.-G.~Mei\ss ner,
  Int.\ J.\ Mod.\ Phys.\ E {\bf 4}  (1995) 193.
  
  \bibitem{Sc02} S.~Scherer,
  Adv..\ Nucl.\ Phys.\ {\bf 27}  (2003) 277.

\bibitem{BM07} V.~Bernard and U.-G.~Mei{\ss}ner,
  Ann.\ Rev.\ Nucl.\ Part.\ Sci.\ {\bf 57} (2007) 33.

\bibitem{Be08} V.~Bernard,
  Prog.\ Part.\ Nucl.\ Phys.\ {\bf 60} (2008) 82.

\bibitem{SS12} 
  S.~Scherer and M.~R.~Schindler,
  \emph{Quantum chromodynamics and chiral symmetry},
  Lect.\ Notes Phys.\  {\bf 830} (2012) 1. (Springer)

\bibitem{Be92} V.~Bernard, N.~Kaiser, J.~Kambor and U.-G.~Mei\ss ner,
  Nucl.\ Phys.\ B {\bf 388} (1992) 315.
  
  \bibitem{Be91} V.~Bernard, N.~Kaiser, and U.-G.~Mei{\ss}ner, Phys. Rev. Lett.
  {\bf 67} (1991) 1515.
 
     \bibitem{Ba97}
    D.~Babusci, G.~Giordano, G.~Matone,
  Phys.\ Rev.\  {\bf C57 } (1998)  291.
  
\bibitem{Fe91} F.~J.~Federspiel, R.~A.~Eisenstein, M.~A.~Lucas {\it et al.},
  Phys.\ Rev.\ Lett.\ {\bf 67} (1991) 1511.

\bibitem{Zi92} A.~Zieger, R.~Van de Vyver, D.~Christmann, A.~De Graeve, C.~Van
  den Abeele and B.~Ziegler,
  Phys.\ Lett.\ B {\bf 278} (1992) 34.

\bibitem{Ha93} E.~L.~Hallin {\it et al.},
  Phys.\ Rev.\ C {\bf 48} (1993) 1497.

\bibitem{MacG95} B.~E.~MacGibbon, G.~Garino, M.~A.~Lucas, A.~M.~Nathan,
  G.~Feldman and B.~Dolbilkin,
  Phys.\ Rev.\ C {\bf 52} (1995) 2097.
  
\bibitem{OdeL} V.~Olmos de Le\'on {\it et al.},
  Eur.\ Phys.\ J.\ A {\bf 10}  (2001) 207.
  
    \bibitem{Mo96}
  C.~Molinari et al.,
  Phys.\ Lett.\  B {\bf 371} (1996) 181.
  
 \bibitem{Pe96}   J.~Peise {\it et al.},
  Phys.\ Lett.\ B {\bf 384} (1996) 37.
  
  \bibitem{Hu97}   A.~H\"{u}nger {\it et al.},
  Nucl.\ Phys.\ A {\bf 620} (1997) 385.

\bibitem{Wi99} F.~Wissmann {\it et al.},
  Nucl.\ Phys.\ A {\bf 660} (1999) 232.
  \bibitem{Wolf} S.~Wolf {\it et al.},
  Eur.\ Phys.\ J.\ A {\bf 12} (2001) 231.

\bibitem{Ca02} M.~Camen {\it et al.},
  Phys.\ Rev.\ C {\bf 65}  (2002) 032202.
  
    \bibitem{Blanpied} G.~Blanpied {\it et al.},
  Phys.\ Rev.\ C {\bf 64}  (2001) 025203
  
  \bibitem{Mi12}
  R.~Miskimen, talk at Conference on Intersections of Particle and Nuclear Physics, St. Petersburg, FL, May 2012, to appear in the proceedings.
  
\bibitem{Mi09}
  R.~Miskimen  et al., HIGS Proposal P-06-09, Measuring the Spin Polarizabilities of the Proton in Double-Polarized Real Compton Scattering (2009).
  
 \bibitem{Ho12}
D.~Hornidge, talk at Conference on Intersections of Particle and Nuclear Physics, St. Petersburg, FL, May 2012, to appear in the proceedings.

  \bibitem{Detmold:2006vu}   W.~Detmold, B.~C.~Tiburzi and A.~Walker-Loud,
  Phys.\ Rev.\ D {\bf 73} (2006) 114505.
  
  \bibitem{De09} W.~Detmold, B.~C.~Tiburzi and A.~Walker-Loud,
Phys.\ Rev.\ D {\bf 79}  (2009) 094505.
  
  \bibitem{De10} W.~Detmold, B.~C.~Tiburzi and A.~Walker-Loud,
Phys.\ Rev.\ D {\bf 81} (2010) 054502.
   
\bibitem{Engelhardt:2011qq} M.~Engelhardt,
  PoS LATTICE {\bf 2011} (2011) 153.

\bibitem{Lujan:2011ue} M.~Lujan, A.~Alexandru and F.~Lee,
  PoS LATTICE {\bf 2011} (2011) 165.

\bibitem{Pachucki} K.~Pachucki, Phys.\ Rev.\ A  {\bf 60} (1999) 3593.

\bibitem{Carlson:2011dz}
  C.~E.~Carlson and M.~Vanderhaeghen,
  arXiv:1109.3779 [physics.atom-ph].  
  
\bibitem{Bi12}   M.~C.~Birse and J.~A.~McGovern,
  Eur.\ Phys.\ J.\ A {\bf 48} (2012) 120.

\bibitem{Ph09} D.~R.~Phillips,
  J.\ Phys.\ G {\bf 36} (2009) 104004.

\bibitem{WalkerLoud:2012bg}
  A.~Walker-Loud, C.~E.~Carlson and G.~A.~Miller,
  Phys.\ Rev.\ Lett.\  {\bf 108} (2012) 232301.

\bibitem{Be93} V.~Bernard, N.~Kaiser, A.~Schmidt and U.-G.~Mei{\ss}ner,
  Phys.\ Lett.\ B {\bf 319} (1993) 269;
  Z. Phys. A {\bf 348}  (1994) 317.
 
\bibitem{Ra93} S.~Ragusa,
  Phys.\ Rev.\ D {\bf 47} (1993) 3757.

\bibitem{McG01} J.~A. McGovern, Phys.\ Rev.\ C {\bf 63} (2001) 064608.

\bibitem{Be02} S.~R. Beane, M.~Malheiro, J.~A. McGovern, D.~R. Phillips, and
  U.~van Kolck,  
  Phys.\ Lett.\ B {\bf 567} (2003) 200
   [Erratum-ibid.\ {\bf 607} (2005) 320].

\bibitem{Be04} S.~R. Beane, M.~Malheiro, J.~A. McGovern, D.~R. Phillips, and
  U.~van Kolck, Nucl.\ Phys.\ A {\bf 747}  (2005) 311.

\bibitem{He96} T.~R.~Hemmert, B.~R.~Holstein and J.~Kambor,
  Phys.\ Lett.\ B {\bf 395} (1997) 89.

\bibitem{He98} T.~R.~Hemmert, B.~R.~Holstein and J.~Kambor,
  J.\ Phys.\ G {\bf 24} (1998) 1831.
  
\bibitem{He97} T.~R.~Hemmert, B.~R.~Holstein, J.~Kambor,
  Phys.\ Rev.\ D {\bf 55} (1997) 5598.
  
\bibitem{Hi04} R.~P. Hildebrandt, H.~W. Grie{\ss}hammer, T.~R. Hemmert and
  B.~Pasquini, Eur.\ Phys.\ J.\ A {\bf 20}  (2004) 293.
  
 \bibitem{PP03} V.~Pascalutsa and D.~R.~Phillips,
  Phys.\ Rev.\ C {\bf 67} (2003) 055202.


\bibitem{BL00} T.~Becher and H.~Leutwyler,
  Eur.\ Phys.\ J.\ C {\bf 9} (1999) 643.

\bibitem{Fu03} T.~Fuchs, J.~Gegelia, G.~Japaridze and S.~Scherer,
  Phys.\ Rev.\ D {\bf 68} (2003) 056005.

\bibitem{LP09} V.~Lensky and V.~Pascalutsa,
  Pisma Zh.\ Eksp.\ Teor.\ Fiz.\ {\bf 89} (2009) 127.
 
  \bibitem{LP10} V.~Lensky and V.~Pascalutsa,
  Eur.\ Phys.\ J.\ C {\bf 65} (2010) 195.
  
\bibitem{Le12}
  V.~Lensky, J.~A.~McGovern, D.~R.~Phillips and V.~Pascalutsa,
  Phys.\ Rev.\ C {\bf 86} (2012) 048201.

\bibitem{McG09} J.~A.~McGovern, H.~W.~Grie\ss hammer, D.~R.~Phillips and
  D.~Shukla,
  PoS CD {\bf 09} (2009) 059.
  
  \bibitem{Ph12} D.~R.~Phillips, J.~A.~McGovern and H.~W.~Grie\ss hammer, talk at Conference on Intersections of Particle and Nuclear Physics, St. Petersburg, FL, May 2012, to appear in the proceedings. [arXiv:1210.3577]
  
  
\bibitem{Ba98}
  D.~Babusci, G.~Giordano, A.~I.~L'vov, G.~Matone and A.~M.~Nathan,
  Phys.\ Rev.\  C {\bf 58} (1998) 1013.

\bibitem{Griesshammer:2001uw}
  H.~W.~Grie\ss hammer and T.~R.~Hemmert,
  Phys.\ Rev.\  C {\bf 65} (2002) 045207.

  \bibitem{FMMS} N.~Fettes, U.-G. Mei\ss ner, M.~Moj\v{z}i\v{s}, and S.
  Steininger,
  Annals Phys.\ {\bf 283} (2000) 273 [Erratum-ibid.\ {\bf 288} (2001) 249].

  \bibitem{MMS} U.-G. Mei\ss ner, G.~M\"uller, and S. Steininger,
  Annals Phys.\ {\bf 279} (2000) 1.
  
\bibitem{Ge99} G.~C.~Gellas, T.~R.~Hemmert, C.~N.~Ktorides and G.~I.~Poulis,
  Phys.\ Rev.\ D {\bf 60} (1999) 054022.
  

\bibitem{hhkk} T.~R.~Hemmert, B.~R.~Holstein, J.~Kambor and G.~Knochlein,
  Phys.\ Rev.\ D {\bf 57} (1998)  5746.

\bibitem{Hildebrandtthesis} R.~P. Hildebrandt, Ph.\ D.\  Thesis, TU M\"unchen
  (2005), nucl-th/0512064.

\bibitem{PV06} V.~Pascalutsa and M.~Vanderhaeghen,
  Phys.\ Rev.\ D {\bf 73} (2006) 034003.

\bibitem{Wa54} K.~M.~Watson,
  Phys.\ Rev.\ {\bf 95} (1954) 228.

\bibitem{LvK09} B.~Long and U.~van Kolck,
  Nucl.\ Phys.\ A {\bf 840} (2010) 39.

\bibitem{Pa10} B.~Pasquini, P.~Pedroni, D.~Drechsel,
  Phys.\ Lett.\ {\bf B687} (2010) 160.

\bibitem{Wissmannmonograph} F. Wissmann, {\it Compton Scattering:
    Investigating the Structure of the Nucleon with Real Photons}, Springer
  Tracts in Modern Physics, Volume 200 (2004).
  
  \bibitem{Baranov:2001} P.~S.~Baranov, A.~I.~Lvov, V.~A.~Petrunkin and N. L.
  Shtarkov,
  Phys.\ Part.\ Nucl.\ {\bf 32} (2001) 376.
  
  \bibitem{PDG} J. Beringer et al. (Particle Data Group), Phys.\ Rev.\ D86 (2012) 010001.
  
  \bibitem{LL00}
  M.~I.~Levchuk and A.~I.~L'vov,
  Nucl.\ Phys.\  A {\bf 674} (2000) 449.
  





\end{thebibliography}
\end{document}